\documentclass[varenna]{cimento}

\usepackage{graphicx}  
\usepackage{color}
\usepackage{ulem}
\usepackage[dvipsnames]{xcolor}
\usepackage{amsmath}
\usepackage{amssymb}
\usepackage[abs]{overpic}

\begin{document}

%

\title{Cosmology: from theory to data, from data to theory}

%
\author{F. Leclercq$^{1,2,3}$, A. Pisani$^{1,2}$ \atque B. D. Wandelt$^{1,2,4}$}
\shortauthor{F. Leclercq, A. Pisani \atque B. D. Wandelt}

%
\institute{$^{1}$ Institut d'Astrophysique de Paris (IAP), UMR 7095, CNRS - UPMC Universit\'e Paris 6, 98bis boulevard Arago, F-75014 Paris, France\\
$^{2}$ Institut Lagrange de Paris (ILP), Sorbonne Universit\'es, 98bis boulevard Arago, F-75014 Paris, France\\
$^{3}$ \'Ecole polytechnique ParisTech, Route de Saclay, F-91128 Palaiseau, France\\
$^{4}$ Departments of Physics and Astronomy, University of Illinois at Urbana-Champaign, Urbana, IL 61801, USA}

\maketitle

\begin{abstract}
Cosmology has come a long way from being based on a small number of observations to being a data-driven precision science. We discuss the questions ``What is observable?", ``What in the Universe is knowable?" and ``What are the fundamental limits to cosmological knowledge?". We then describe the methodology for investigation: theoretical hypotheses are used to model, predict and anticipate results; data is used to infer theory. We illustrate with concrete examples of principled analysis approaches from the study of cosmic microwave background anisotropies and surveys of large-scale structure, culminating in a summary of the highest precision probe to date of the physical origin of cosmic structures: the Planck 2013 constraints on primordial non-Gaussianity.

\vspace{8 mm}

\textit{Proceedings of the International School of Physics ``Enrico Fermi" of the Italian Physical Society – SIF-Course CLXXXVI: ``New Horizons for Observational Cosmology", June 30-July 6, 2013, Varenna, Italy}
\end{abstract}

%

\tableofcontents
\section*{Introduction: the big picture}

Physical cosmology is a science based on high-precision observations. In the modern era, large surveys are designed to address questions that might seem immodest, but certainly enthralling: ``How did the Universe begin?'' (if it did!), ``How did structure appear in the Universe?", ``How did it evolve until today?", ``What is the Universe made of?", ``What are the properties of the dominants components of the Universe, known by the placeholder terms `dark matter' and `dark energy'?" ``What is the geometry and the symmetry of the Universe?". All cosmological observations are informative in some ways about these questions, but the message is encoded and sometimes hard to extricate.

\begin{figure}
\begin{center}
\includegraphics[width=\textwidth]{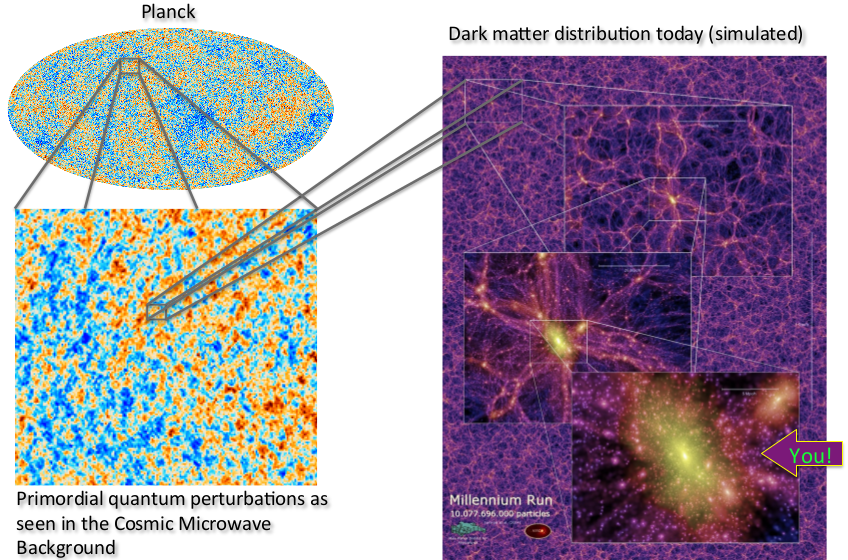}
\caption{The big picture in cosmology. Primordial perturbations as seen in the Cosmic Microwave Background anisotropies (Planck), on the left, forming the seeds for the dark matter distribution today (Millenium Run simulation), on the right. Cosmostatistics predicts observables from primordial random inputs and uses the stochastic departures from homogeneity on all observable scales to distinguish between cosmological models.\label{fig:big_picture}}
\end{center}
\end{figure}

The cosmic microwave background (CMB) gives an image of the primordial perturbations: it is a screen-shot of the past, about 380,000 years after the Big Bang. Present-day galaxies are not just randomly distributed, but trace the underlying cosmological mass distribution: there exists a large-scale structure (LSS) in the Universe, which is seeded by the same primordial perturbations (see fig. \ref{fig:big_picture}). In the simplest models of the inflationary paradigm, the initial conditions, of quantum origin, are very nearly Gaussian. In this picture, all the information from the beginning of the Universe forms a single Gaussian random field.

We will refer to the discipline that deals with stochastic quantities as seeds of structure in the Universe as \textit{cosmostatistics}. It consists of predicting cosmological observables from random inputs (\textit{from theory to data}) and uses the departures from homogeneity and isotropy, observed in astronomical surveys, to distinguish between cosmological models (\textit{from data to theory}). The continuous exchange between data and theory is what allows us to make progress as we consider the immodest questions of cosmology. The essential point of these lectures is to address the functional aspects of this interplay.

Before we dive into these more technical issues, we pause to reflect on the nature of cosmological data. The precision measurement of the CMB anisotropies provided by the Planck satellite is a recent major milestone of cosmology. But is the CMB the ultimate cosmological probe? Is there more to know about the Universe? Are we looking at all the information we could have access to? Are there fundamental limits to cosmological information? What is \textit{observable} in the Universe? 

This document is structured as follows. In section \ref{sec:Observable and knowable}, we present known sources of cosmological information whether already exploited or envisioned for the future, and the associated physical phenomena. We discuss causal diagrams and show they provide quick and correct answers about what is observable and knowable in the Universe. In section \ref{sec:Primordial perturbations and the CMB}, we review standard results about cosmological perturbations: their birth and their analysis in terms of a Gaussian random field. In section \ref{sec:Bayesian cosmostatistics}, we present statistical methods used to treat cosmological models beyond Gaussianity, in a Bayesian framework. Finally, section \ref{sec:Applications of inference} illustrates two applications of inference: the reconstruction of initial conditions from large-scale structure surveys and the Planck 2013 constraints on primordial non-Gaussianity.
\section{Observable and knowable}
\label{sec:Observable and knowable}

This section provides an overview of what is ultimately \textit{observable} and \textit{knowable} in the Universe. In $\S$ \ref{sec:Cosmological probes}, we present the astrophysical probes, thought to contain all the information available for cosmologists. In $\S$ \ref{sec:Limits to cosmological knowledge, causal diagrams}, we examine the fundamental limits to the information we can access via these probes, due to the causal structure of the Universe. In doing so, we discuss causal diagrams, a convenient way to represent the information accessible directly and indirectly, at different moments.

\subsection{Cosmological probes}
\label{sec:Cosmological probes}

\begin{figure}
\begin{center}
\includegraphics[width=0.85\textwidth]{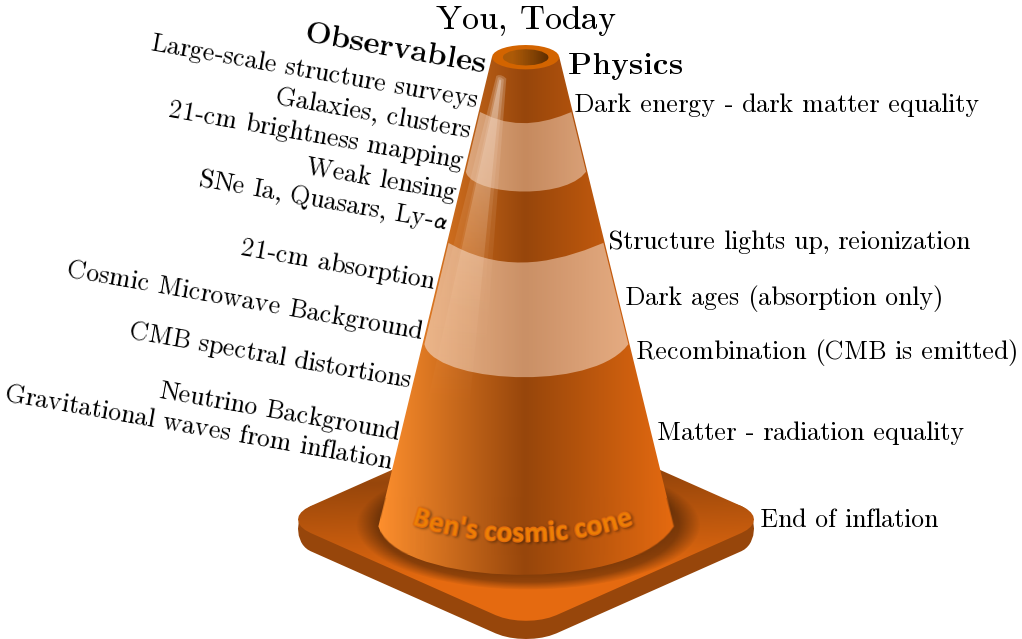}
\caption{Ben's cosmic cone: schematic representation of the relativistic light cone from a cosmologist's point of view. On the left, the cosmological observables, already observed or predicted. On the right, the physical phenomena they relate to, in the standard cosmological model.\label{fig:Ben_cone}}
\end{center}
\end{figure}

In order to catch a glimpse of the wealth and diversity of available cosmological information, it is convenient to place on a relativistic light cone both the observables and the related physical phenomena in the history of the Universe (see fig. \ref{fig:Ben_cone}). The probes cover a wide range of redshifts, from the inflationary phase to the present-day Universe. Below, we give a broad panorama in the form of a succinct description of each of them, from the youngest to the oldest.

\begin{itemize}
\item \textit{The large-scale structure of the Universe} ($0 \leq z \lesssim 2$). The distribution of structures at the largest scales is not random, but forms a cosmic web, composed of voids, filaments, walls and clusters. Galaxies trace this structure, the detailed appearance of which retains a memory of its formation history.

\item \textit{21 cm spectral line brightness mapping} ($0 \leq z \lesssim 3$). A quantum transition in the hyperfine structure of the hydrogen atom (essentially, the reversal of the electron's spin) is responsible for the emission of a characteristic spectral line with a vacuum wavelength of 21 cm. This emission line carries information about the structure in the Universe, since neutral hydrogen traces the distribution of galaxies and dark matter. The 21 cm line is widely observed in radio astronomy. Systematic cosmological surveys are also planned; they are of particular interest because they allow a very precise measurement of redshift, since the emission mechanism is well understood.

\item \textit{Type Ia supernovae} ($0 \leq z \lesssim 3$). Their use as \textit{standard candles} allows a determination of cosmological distances \cite{Riess1998, Perlmutter1999}, therefore probing the late-time expansion history of the Universe. In particular, they are sensitive to the properties of dark energy.

\item \textit{Weak gravitational lensing} ($0.05 \lesssim z \lesssim 3$). The general-relativistic deflection of light-rays by matter allows a statistical reconstruction of the gravitational potential in the sky. The method is sensible to the mass distribution independently of its composition or dynamical state, which means that it probes the distribution not only of visible matter, but also of dark matter.

\item \textit{Quasars} ($0.05 \lesssim z \lesssim 7$). The absorption of the redshifted Lyman-$\alpha$ spectral line in the spectra of quasars (so-called \textit{Lyman-$\alpha$ forest}) is used as a probe of the properties of the intergalactic medium and is expected to yield estimates of cosmological parameters ($H_0$, $\Omega_\mathrm{m}$, $\Omega_\Lambda$).

\item \textit{21 cm absorption} ($8 \lesssim z \lesssim 1000$). The absorption of the 21 cm spectral line of atomic hydrogen (see above) is of particular interest because it is the only known way of probing the cosmic ``dark ages" from recombination to reionization. The detailed observation of the absorption of the 21 cm background is expected to provide a picture of how the Universe was reionized.

\item \textit{The cosmic microwave background} ($z = 1089$). It consists of the photons emitted during the time of recombination of electrons onto nuclei and simultaneous decoupling of radiation from matter. The temperature anisotropies of the CMB, measured to great accuracy by the Planck satellite, are one of the most famous cosmological observables. They allow a precise determination of cosmological parameters and are a powerful probe of the early Universe. In the near future, the polarization pattern of CMB photons (divided in curl-free components called E-modes, and rotational components called B-modes) is expected to provide information about inflationary gravitational waves (see below), as well as on the intervening density perturbations, via weak gravitational lensing (cosmic shear).

\item \textit{Baryon Acoustic Oscillations} ($z \gtrsim 1100$). The baryon acoustic oscillations (BAO) refer to periodic fluctuations in the density of matter in the Universe, caused by acoustic waves in the primordial photon-baryon plasma. This early universe phenomenon gets imprinted in particular in the CMB and in the LSS. BAO matter clustering provides a \textit{standard ruler} for length scales in cosmology from which it is possible to probe the expansion history and extract cosmological information, in particular about dark energy.

\item \textit{Isocurvature perturbations} ($z \gtrsim 1100$). In addition to the so-called adiabatic modes (fluctuations of the overall local matter density), there may exist perturbations in the particle density ratio between two fluids (cold dark matter and radiation, baryons and radiation, etc.). Primordial isocurvature perturbations would reveal the existence of several degrees of freedom during inflation (additional dynamical fields or simply spectator fields). Since they leave distinctive features in the CMB anisotropies and in the LSS, they can in principle be disentangled from the usual adiabatic modes.

\item \textit{CMB $\mu$-distortions} ($10^5 \lesssim z \lesssim 10^8$) \textit{and $y$-distortions} ($1000 \lesssim z \lesssim 10^5$). The photon-baryon thermodynamic equilibrium between big-bang nucleosynthesis and recombination can be slightly perturbed by different physical phenomena. This leads to spectral distortions of the CMB: it is not a perfect black body. Early-time $\mu$-distortions are present if a small chemical potential exists and late-time $y$-distortions are due to Compton scattering of photons on electrons (equivalent to the thermal Sunyaev-Zel'dovich effect). These probes of early Universe physics are the object of mission proposals involving a very sensitive spectrometer to measure deviations of the CMB spectrum from a perfect black body spectrum.

\item \textit{The cosmic neutrino background} ($z \simeq 10^9$). In the Hot Big Bang model, neutrinos decouple from the rest of the primordial plasma and should form a neutrino background (CNB) in a similar way as CMB photons do. Due to their weak interactions with matter, neutrinos decouple much earlier than photons (at a temperature $T_\nu \simeq 1$ MeV compared to $T_\gamma \simeq 0.3$ eV). For this reason, the CNB would have a much higher number of super-horizon modes than the CMB, probe larger scales and a much younger Universe. The detection of such a background would be a major triumph for the standard Hot Big Bang cosmological model. Unfortunately, despite the high density of cosmological neutrinos, a direct detection is extremely difficult due to their very low energy and cross-section.

\item \textit{Gravitational waves from inflation} ($z \gtrsim 10^{25}$). Gravitational waves are the excitations of the tensor modes of the metric. A stochastic background of primordial gravitational waves is a generic prediction of inflation. It has not been detected yet, but could be observed thanks to spatial or ground-based interferometers (although astrophysical sources are expected to be seen much sooner) or through B-modes in the CMB polarization sourced by gravitational wave shear at decoupling.

\end{itemize}

\subsection{Limits to cosmological knowledge, causal diagrams}
\label{sec:Limits to cosmological knowledge, causal diagrams}

What can we extract from these cosmological observables? What information about the Universe is accessible to us, directly or indirectly? Are there fundamental limits to cosmological knowledge?

Generally, the intrinsic limits to the information we can have access to are due to its finite speed of propagation asserted by special relativity. For this reason, there exists a causal structure of the Universe that is relevant for cosmology. In particular, it is only possible to observe part of the Universe at a given time. This fact limits the information available for making statistical statements about scales comparable to the entire observable Universe. Since we only have access to a single realization from the ensemble of universes that could have arisen, statements about the largest scales are subject to uncertainty, usually referred to as \textit{cosmic variance}.

Causal diagrams are a convenient tool to visualize the information accessible directly or indirectly. They depict relativistic light cones, the surfaces describing the temporal evolution of light rays in space-time. These include both a future part (everything that you can possibly influence) and a past part (everything that can possibly have influenced you). On causal diagrams, your world line, i.e. your trajectory in space-time, is essentially a straight line at the spatial origin (the $t$-axis). As usual, for convenience in graphical representations, we will suppress one spatial dimension and represent the four-dimensional space-time in 2+1 dimensions. In addition, we will use comoving coordinates to factor out the expansion of the Universe, so that light-rays travel on diagonal lines\footnote{We present a simplified discussion that ignores the effect of dark energy domination on the future -- for a more complete treatment see \cite{Loeb2012}.}.

To delve into the causal structure of our Universe, we will successively consider three categories: the information we can access now, directly; the information we could access directly, in a Universe's lifetime and the information we can access indirectly. 

\subsubsection{Information accessible directly, now}

\begin{figure}
\begin{center}
\includegraphics[width=0.6\textwidth]{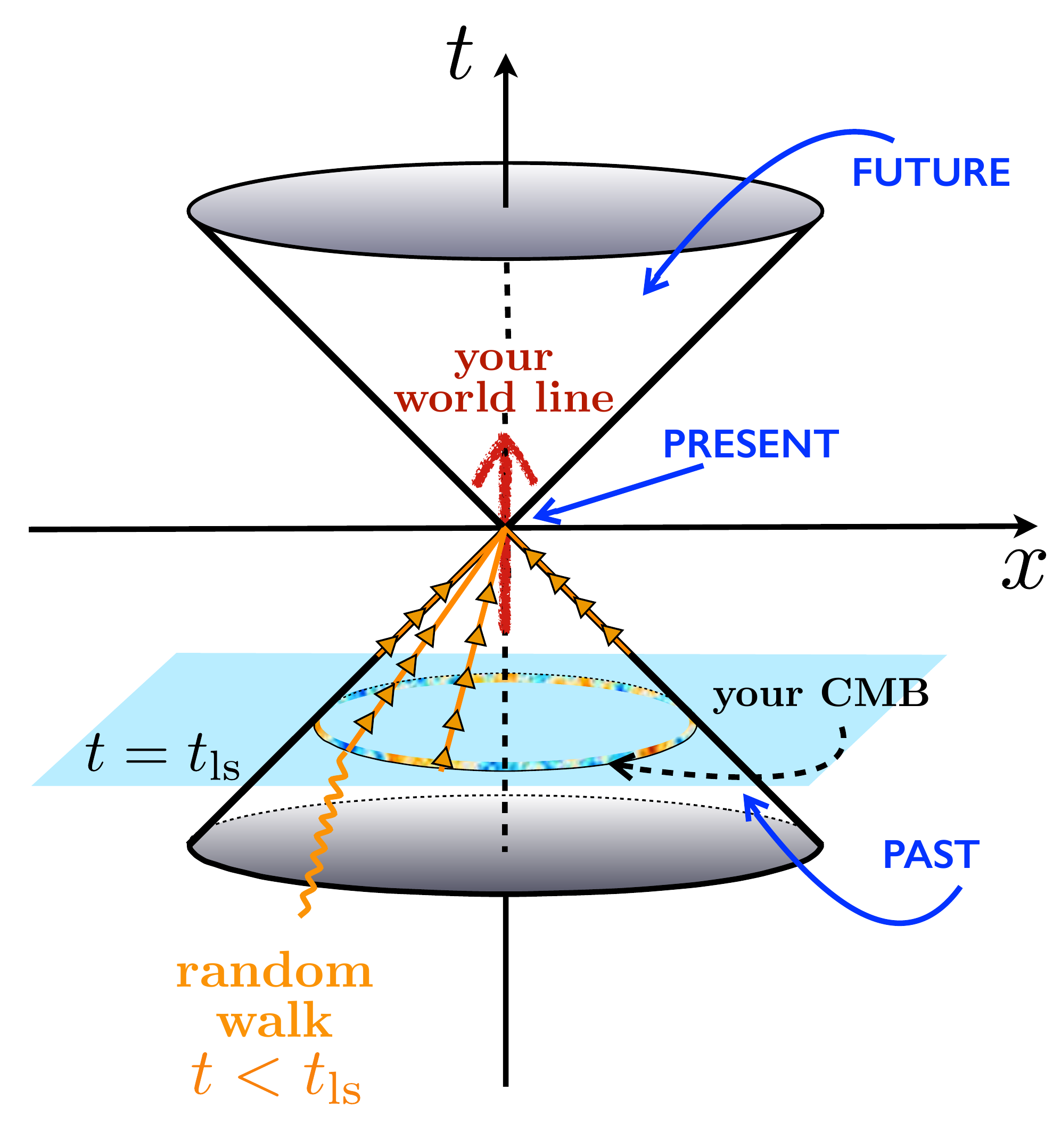}
\end{center}
\caption{\textit{Direct knowledge light cone}: the interior of your past light cone (4D volume) shows everything that could possibly have influenced you; the upper part of the cone shows everything that you can possibly influence. The red arrow represents your world line. The sky blue plane shows the time of last scattering. Its intersection with your light cone is your ``CMB circle". The yellow triangles show examples of the path of photons that reach you now. Before decoupling, they follow a random walk with increasing mean free path in the interior of your light cone. After decoupling, they travel on the surface of the cone (3D volume).\label{fig:direct_now}}
\end{figure}

-- Causality allows direct access, now, to:
\begin{itemize}
\item the surface of your past light cone (a 3-dimensional volume): all the photons that reach you now (e.g. photons from distant galaxies or from the CMB),
\item the interior of your past light cone (a 4-dimensional volume): all events that could possibly influence you via a slower-than-light signal (this includes all massive particles that you receive from space).
\end{itemize}

Figure \ref{fig:direct_now} shows your light cone and the information you can have access to. Your ``CMB circle" (the last-scattering sphere in 3D) is the intersection of a plane (corresponding to the time of last scattering $t=t_\mathrm{ls}$ i.e. the time when the CMB was emitted) and your past light cone.

\subsubsection{Information accessible directly, over time}

\begin{figure}
\begin{center}
\includegraphics[width=0.9\textwidth]{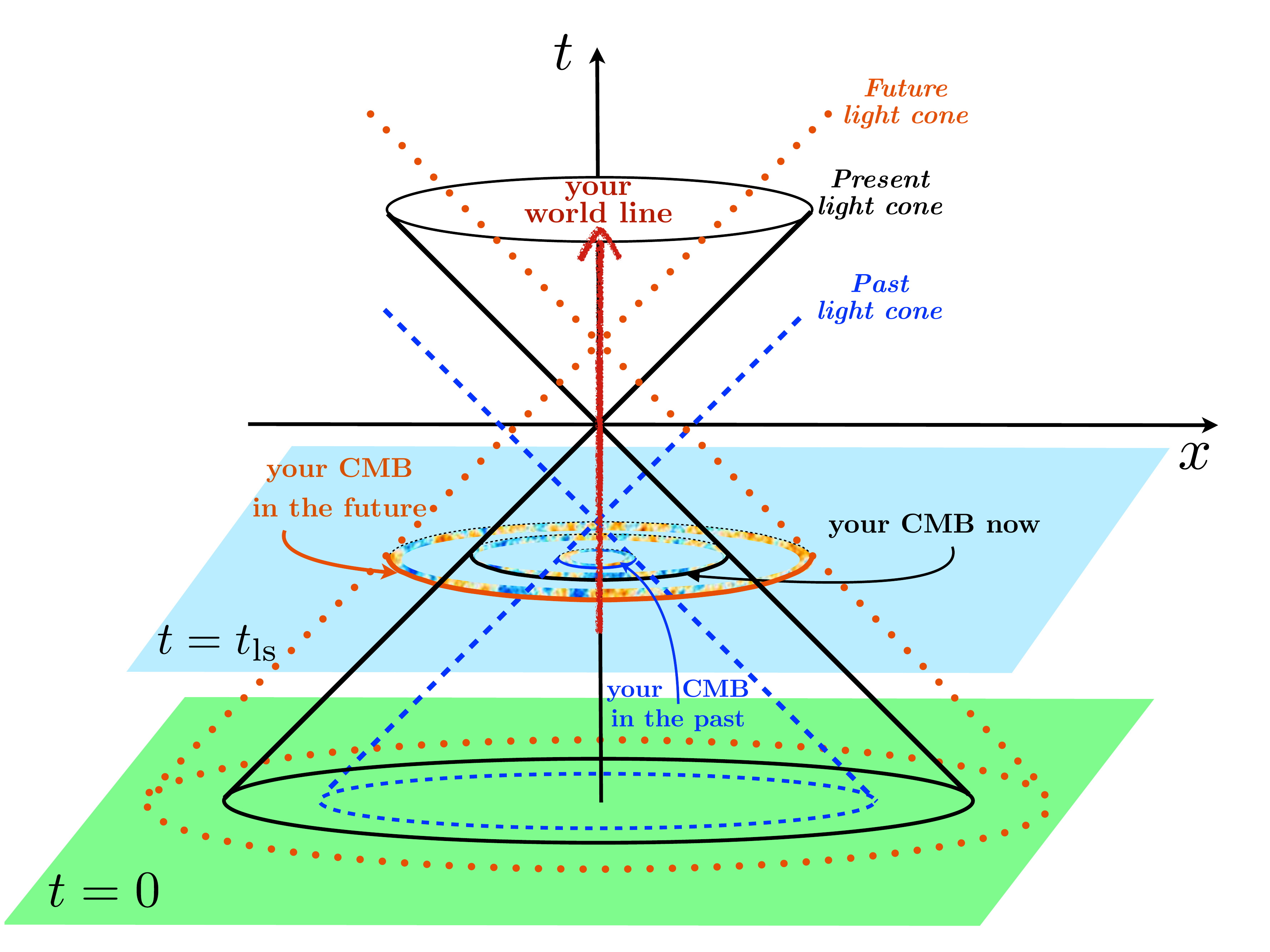}
\end{center}
\caption{\textit{Whole life light cone}: the volume occupied by a light cone at every point of your world line corresponds to everything that you have ever seen and will ever see. Here we only show a light cone for the past (dashed blue line) and one for the future (dotted orange line). All light cones intersect differently the time of last scattering surface (sky blue plane), meaning that your CMB circle is time-dependent. Note also, that if the age of the Universe is finite, at any given time, there exist regions of the $t=0$ plane that are not causally connected to you.\label{fig:whole_life}}
\end{figure}

-- If you want to have access to more information, what about just waiting (or starting your cosmological observations earlier)? At each moment of your world line there is a light cone. As you move along your world line these light cones sweep out a 4D volume that includes everything you have ever seen and will ever see (see fig. \ref{fig:whole_life}). This is a bigger 4D volume than previously considered. However, note that if the age of the Universe is finite (or if there is an event horizon \cite{Loeb2012}), at any given time, there are regions of the $t=0$ plane that you have not yet seen.

The CMB you have access to changes with time, because the intersection of the plane corresponding to the time of last scattering $t=t_\mathrm{ls}$ and the light cone changes when you consider a different light cone. This means that in principle, waiting (for a long time!) allows access to a thick ring in the last scattering plane, i.e. turns the CMB into a three-dimensional map.

\subsubsection{Information accessible indirectly}

\begin{figure}
\begin{center}
\includegraphics[width=0.8\textwidth]{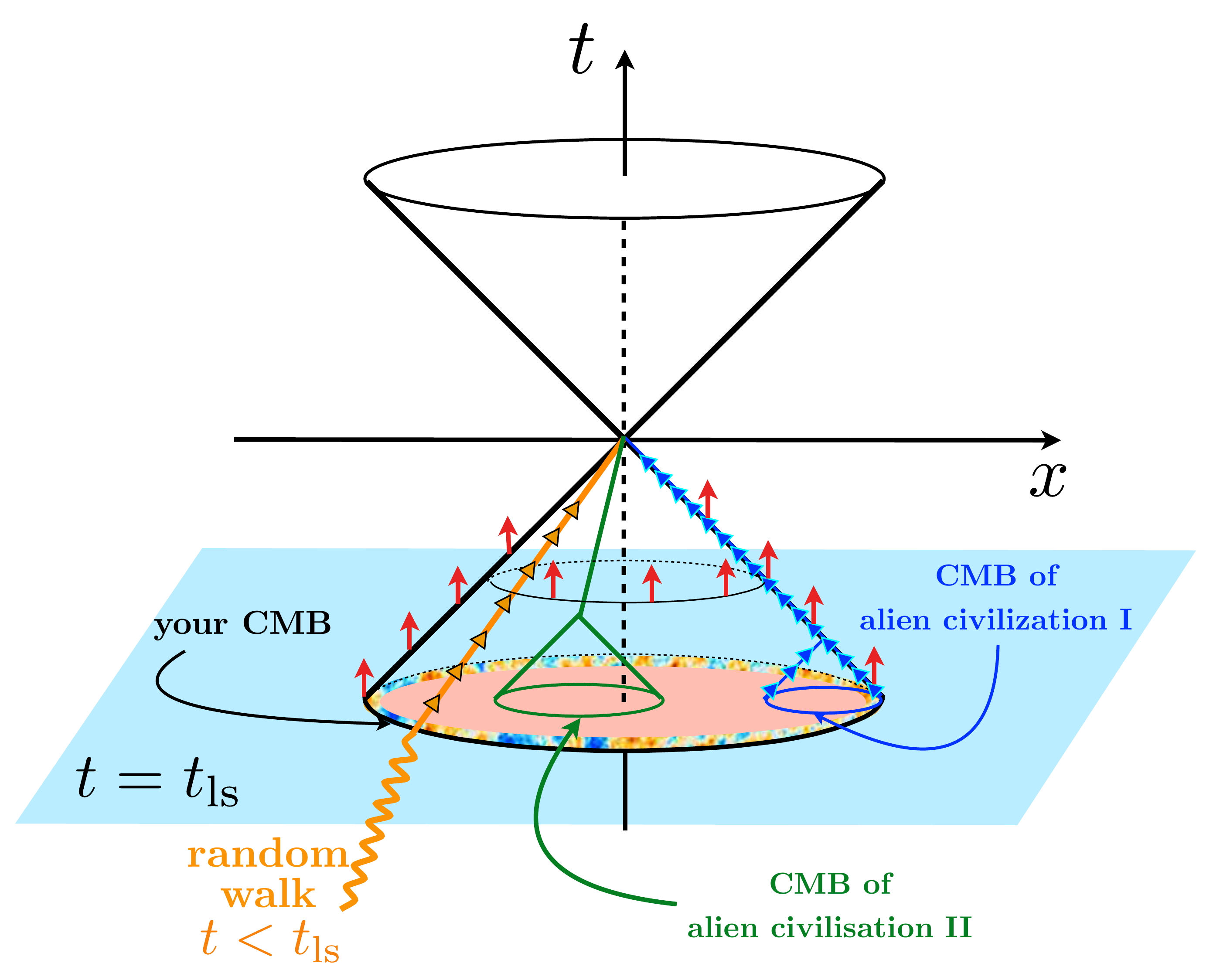}
\end{center}
\caption{\textit{Indirect knowledge light cone}: knowing the laws of physics, you can evolve forward your observations and make predictions for events outside of your light cone (red arrows). If a friendly alien civilization had observed the CMB in the past and sent the map by radio transmission (along the surface of your light cone), we would have access to a new CMB circle tangent to ours (in blue). With the additional help of a time-capsule or of a second alien civilization traveling in a slower-than-light spacecraft, we would get another CMB circle in the interior of our light cone (in green). All this additional information is a way to fight cosmic variance.\label{fig:indirect_knowledge}}
\end{figure}

-- Suppose now, that you want more information than directly accessible at the moment, but do not have time to wait. Then, the only way to proceed is indirectly. 

Knowing the laws of physics, you can infer the behaviour of the Universe in different regions of space-time, from what you see right now. In particular, you can evolve observations forward or backward and predict events in the interior of your light cone (evolution backward) or outside your light cone (evolution forward). Unfortunately, these predictions cannot be tested (at least, directly and now).

As a provocative example, imagine the existence of friendly aliens that observed the CMB in the past and sent it to us by radio transmission (on the surface of our light cone, see fig. \ref{fig:indirect_knowledge}). This CMB circle is not the same as ours! We would get a new (smaller) CMB circle in the interior of ours (tangent to ours in this case). Analogously, with the help of an alien spacecraft (slower than light) or a time capsule (a durable container, which protects a message or objects intended for future generations) we could also obtain a new CMB circle in the strict interior of our light cone. With a lot of these circles, we could fill the whole disk and have access to a 3D CMB map with many more modes and therefore much more information, which would be a way to do better with cosmic variance than if we only had access to ``our'' CMB.

Are these thought experiments just science-fiction, or actually useful for physical cosmology? Several realistic effects act as ``friendly aliens-like" scenarii. In particular, it is worth mentioning:

\begin{itemize}
\item \textit{The polarized Sunyaev-Zel'dovich effect} \cite{Kamionkowski1997, Cooray2003}. Charged particles in pockets of hot gas inside clusters can scatter CMB photons before they reach us. Since Compton scattering is angle-sensitive, if the input has a quadrupole anisotropy, the output is polarized. Thereby, polarization of the CMB probes the quadrupole felt by photons through the Sunyaev-Zel'dovich effect. Since each cluster on the line of sight has its own light cone, a measurement of this effect would amount to measuring the quadrupoles of each cluster's CMB sky.
\item \textit{Inhomogeneous reionization} \cite{Dore2007}. In this scenario, reionization of the Universe does not happen everywhere at the same time at the end of the dark ages, but instead starts in different pockets of hot gas inhomogeneously distributed. A statistical treatment of these patches allows to extract cosmological information in the interior of our light cone.
\item \textit{Some other ``time capsules"}: the abundance of chemical elements or isotopes thought to be primordial (e.g. primordial deuterium at Big Bang nucleosynthesis, three minutes after the Big Bang); dark matter particles in the galactic halo; neutrinos (since they are massive, they move slightly in the interior of our light cone);  or cosmic rays (though they tend to move along the light cone in the absence of magnetic fields, since they are very energetic).
\end{itemize}
\section{Primordial perturbations and the CMB}
\label{sec:Primordial perturbations and the CMB}

\subsection{The birth of perturbations}
\label{sec:The birth of perturbations}

We will consider the Hot Big Bang scenario and inflation as an observationally well-supported physical model for the initial conditions \footnote{The following discussion of cosmic inflation and its observational consequences is deliberately quite ``canonical'', in that (mostly) it cheerfully ignores  the ongoing debate regarding the important, and currently unresolved quantum-cosmological puzzles that are part and parcel of the inflationary paradigm. Going beyond the standard description opens up a (very interesting) can of worms, a careful presentation of which would take us too far away from the main thread of these lectures. Interested readers are invited to study the inflationary ``unlikeliness problem'' as described by Ijjas, Steinhardt \& Loeb \cite{Ijjas2013} and the answer of Guth, Kaiser \& Nomura \cite{Guth2013}, an analysis of the relative likelihood of inflation with respect to other scenarii \cite{Albrecht2004}, a description of the breakdown of the Born rule for multiple identical observers in an inflationary landscape \cite{Page2009}, and a review on the measure problem in cosmology \cite{Gibbons2008}.}. The inflationary paradigm provides explanations for some shortcomings of the standard Hot Big Bang picture, such as the horizon problem. According to this picture, during the inflationary era, the equation of state of the Universe is governed by a potential-dominated quantum scalar field with negative pressure, the so-called \textit{inflaton field}. This quantum field drives an exponential growth of the cosmic scale factor. What is remarkable with inflation is that the accelerated expansion in the very early Universe can magnify the vacuum quantum fluctuations of the inflaton into macroscopic cosmological perturbations.

An intuitive sketch of the quantum origin of density perturbations is as follows:
\begin{itemize}
\item The vacuum expectation value of the inflaton field is slightly spread by quantum fluctuations: $\delta \phi(\mathbf{x})$...
\item ... which induces a local time delay for the end of inflation, $\delta t(\mathbf{x})$...
\item ... which translates into density fluctuations after inflation, $\delta \rho(\mathbf{x})$...
\item ... which become the CMB anisotropies $\delta T(\mathbf{x})$, the inhomogeneous galaxy distribution, $\delta n_\mathrm{g}(\mathbf{x})$, etc.
\end{itemize}

Two very good introductory reviews on the subject are \cite{Baumann2011,Langlois2010}. For concreteness here, we will consider the simplest model for inflation (a single scalar field $\phi$ minimally coupled to gravity, slowly rolling down his potential), for which the action takes the following form:
\begin{equation}
S = \int \mathrm{d}^4 x \, \sqrt{-g} \left[ \frac{M_{\mathrm{Pl}}^2}{2}\mathcal{R} - \frac{1}{2} g^{\mu\nu} \partial_{\mu} \phi \, \partial_{\nu} \phi - V(\phi) \right] .
\end{equation}

In the comoving gauge (defined in slow-roll inflation by the vanishing of the inflaton perturbation, $\delta \phi = 0$), perturbations are characterized purely by metric fluctuations,
\begin{equation}
\delta g_{ij} = a^2(1-2\zeta) \delta_{ij} + a^2 h_{ij} .
\end{equation}

\begin{figure}
\begin{center}
\includegraphics[width=0.4\textwidth]{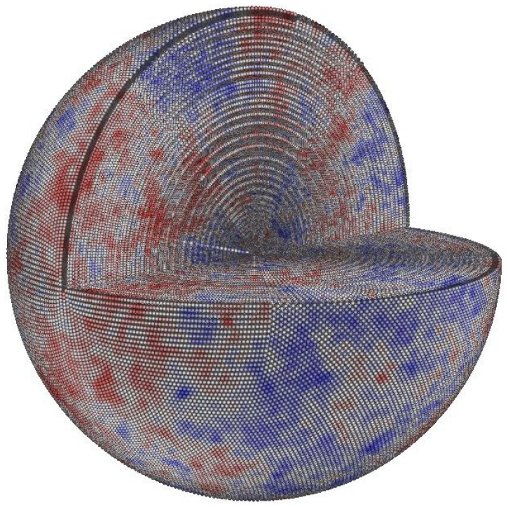} 
\end{center}
\caption{Example for simulated curvature perturbations visualized on the light cone. The linear gravitational potential (proportional to the comoving curvature perturbations) is shown on different shells from the observer (center) to the last scattering surface (outermost shell). Figure adapted from \cite{Elsner2009}.\label{fig:potential}}
\end{figure}

\begin{figure}
\begin{center}
\includegraphics[width=0.9\textwidth]{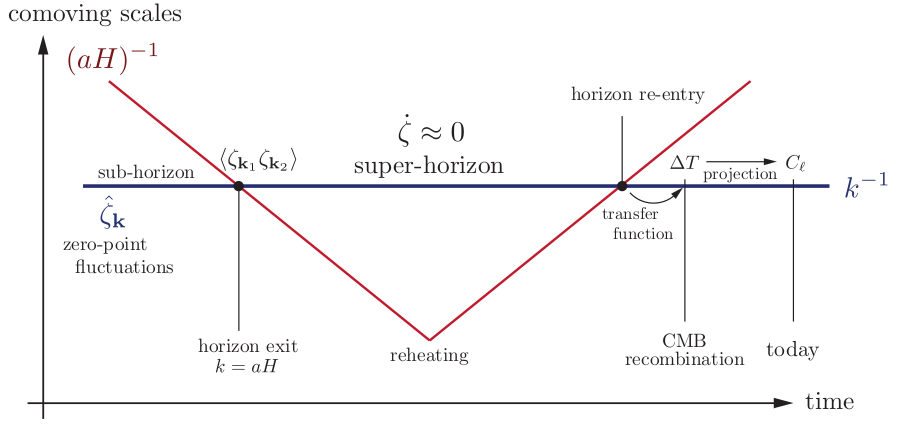} 
\end{center}
\caption{Curvature perturbation during and after inflation. The comoving Hubble radius $(aH)^{-1}$ shrinks during inflation and grows in the Hot Big Bang phase. This implies that comoving scales of wavenumber $k$ exit the horizon at early time and re-enter the horizon at a later time (Hubble-crossing). The comoving curvature perturbations are conserved on superhorizon scales. This allows us to be ignorant of the uncertain details of the reheating phase when relating the curvature perturbation at horizon exit during the early inflationary phase and the late-time observables. Figure from \cite{Baumann2011}.\label{fig:inflation}}
\end{figure}

The comoving curvature perturbation $\zeta$, represented in fig. \ref{fig:potential}, has the crucial property of being time-independent on superhorizon scales (for adiabatic matter fluctuations, see fig. \ref{fig:inflation}). The general calculation of quantum fluctuations during inflation shows that cosmological fluctuations are a combination of gauge-invariant perturbations of the metric and of the inflaton field, described by the canonically-normalized Mukhanov variable, $v~\equiv~z\zeta$, where $z~\equiv~a \phi'/\mathcal{H}$, $\mathcal{H}=a'/a$ is the conformal Hubble parameter, and a prime denotes a derivative with respect to conformal time. The Fourier modes of $v$, denoted by $v_{\mathbf{k}}$, follow a parametric amplifying equation of motion (the Mukhanov-Sasaki equation):
\begin{equation}
v_{\mathbf{k}}'' + \left( k^2 - \frac{z''}{z} \right) v_{\mathbf{k}} = 0 .
\end{equation}

The power spectrum for the fluctuations of $v$ is found to be $\mathcal{P}_v = (aH)^2/(2k^3)$, which sources both scalar curvature perturbations, with a power spectrum $\mathcal{P}_\zeta = \mathcal{P}_v/z^2$ and tensor perturbations (a stochastic background of gravitational waves), with a power spectrum $\mathcal{P}_\mathrm{t} = 2 \times (2/a M_\mathrm{Pl})^2 \, \mathcal{P}_v$. We define the power spectra in dimensionless form by $\Delta_\mathrm{s}^2(k) \equiv k^3/(2\pi^2) \, \mathcal{P}_\zeta$ and $\Delta_\mathrm{t}^2(k) \equiv k^3/(2\pi^2) \, \mathcal{P}_\mathrm{t}$, and the tensor-to-scalar ratio by $r \equiv \Delta_\mathrm{s}^2/\Delta_\mathrm{t}^2$. Since $(aH)^2$ is a function of time, the resulting Universe will deviate slightly from a Harrison-Zel'dovich spectrum ($r = 0$, i.e. no gravitational waves, and a strictly scale-invariant scalar power spectrum, i.e. $\Delta_\mathrm{s}^2 (k) = \mathrm{const.}$). The usual way to quantify the deviations from scale-invariance is via the scalar and tensor spectral indices, 
\begin{equation}
n_\mathrm{s} - 1 \equiv \frac{\mathrm{d}\ln \Delta_\mathrm{s}^2}{\mathrm{d}\ln k} \quad \mathrm{and} \quad n_\mathrm{t} \equiv \frac{\mathrm{d}\ln \Delta_\mathrm{t}^2}{\mathrm{d}\ln k} .
\end{equation}

In slow-roll inflation, $n_\mathrm{s}$, $n_\mathrm{t}$ and $r$ are all linked to the slow-roll parameters, which yields the following \textit{consistency relation}, relating purely observable quantities:
\begin{equation}
r = -8 n_\mathrm{t} .
\end{equation}

In the uniform density gauge, where $\delta \phi = 0$, and for super-horizon scales, the primordial gravitational potential sourcing the density perturbations is simply $\Phi = -\zeta$. Therefore, at the end of inflation, this model naturally provides us with a statistically homogeneous and isotropic density field with small, very nearly Gaussian-distributed, and nearly scale-invariant density perturbations.

The anisotropies in the CMB temperature and polarization are -- to a very good approximation -- linear maps of the initial perturbations. Even for the present-day galaxy distribution, smoothing out small scale power (affected by non-linear gravitational evolution) yields a nearly Gaussian random field on the largest scales. For this reason, one has to describe the linear transport from one field to another. For the CMB, linear radiative transfer relates the primordial perturbations to the temperature anisotropies:
\begin{equation}
a_{\ell m} = \int \mathrm{d}^3 \mathbf{k} \, \zeta(\mathbf{k}) \, g_{\ell m}(\mathbf{k}) ,
\label{eq:transfer_function}
\end{equation}
where $a_{\ell m}$ are the components of the temperature fluctuations $\delta T/T$, decomposed into spherical harmonics, and $g_{\ell m}$ is the transfer function describing all the linear physical processes of interest. For a homogeneous and isotropic universe, the transfer function takes a simple form,
\begin{equation}
g_{\ell m}(\mathbf{k}) = g_\ell(\left| \mathbf{k} \right|) \, i^\ell \, Y_{\ell m}^{*} (\mathbf{\hat{k}}) .
\end{equation}

A period of inflation in the early universe explains why the universe is homogeneous, isotropic, and flat. Furthermore, all predictions concerning the statistics of primordial perturbations (originally seen as a byproduct of inflation) are broadly compatible with all observations so far, including the Planck 2013 results. Phenomenologically, inflation is therefore a great success. But what is the physics of inflation, i.e. what microphysical phenomena lead to an accelerated expansion? At this point, it is necessary to stress that inflation is not a specific model, but rather a paradigm encompassing a wide class of models. Since the CMB gives an image of the early Universe, it can help us discriminate among inflationary models. Inflation is thus an example of the interplay between theory and data: the concept is designed to explain observational facts, and is then tested back by data. 

How do we use the CMB to constrain inflation? Before discussing primordial non-Gaussianity ($\S$ \ref{sec:Planck results on primordial non-Gaussianity}), the simplest solution to this problem involves describing the CMB and the initial perturbations as Gaussian random fields ($\S$ \ref{sec:Gaussian random fields}) and inverting the linear radiative transfer relating them ($\S$ \ref{sec:The linear physics CMB time-machine}).

\subsection{Gaussian random fields}
\label{sec:Gaussian random fields}

This section summarizes some results about Gaussian random fields (GRFs) with emphasis on the properties useful for the analysis of the CMB. For a more general review of GRFs in cosmostatistics, see \cite{Wandelt2013}. 

\subsubsection{Definition}

-- A $n$-dimensional vector $x$ is a Gaussian random field (we will often say ``is Gaussian" in the following) with mean $\mu$ and covariance $C$ if it has the following probability density function (pdf)
\begin{equation}
p(x|\mu,C)= \frac{e^{-\frac12 (x-\mu)^TC^{-1}(x-\mu)}}{\sqrt{|2\pi C|}}.
\end{equation}

As can be seen from the definition, a GRF is completely specified by its mean $\mu$ and its variance $C$. It is easy to check that the mean is really $\left< x \right> = \mu$ and the covariance is really $\left< (x-\mu)(x-\mu)^T \right> = C$ by just calculating the Gaussian integrals. 

There are many software packages that allow generating single Gaussian random variates with mean 0 and variance 1, i.e. \textit{normal variates}, e.g. using the well-known Box-M\"uller method. Using a $n$-vector $\xi$ of such normal variates we can generate random realizations of a GRF with covariance $C$ and mean $\mu$ by simply taking any matrix $\sqrt{C}$ that satisfies $\sqrt{C}\sqrt{C}^{T}=C$ and computing $x=\sqrt{C}\xi+\mu$. One general way to generate $\sqrt{C}$ under the condition that $C$ has only positive definite eigenvalues is to use the so-called Cholesky decomposition, implemented in many numerical packages. It is easy to verify that $x$ has the right mean and covariance using that $\left< \xi \xi^{T}\right>=\textbf{1}$.

\subsubsection{Moments of Gaussian random fields and Wick's theorem}

-- So we can calculate $\left< x \right>$ and $\left< xx^T \right>$. What about higher order moments? Let us focus on central moments e.g. $\left< (x-\mu)(x-\mu)^T \right> = C$, since it is always easy to put the mean back in. Equivalently, we look at the moments for $\mu=0$ (we will assume $\mu=0$ from now on). We will also put back the explicit indices on the vectors.

Any odd (central) moments, e.g. the third ($\left<x_{i}x_{j}x_{k}\right>$), fifth ($\left<x_{i}x_{j}x_{k}x_{l}x_{m}\right>$) etc., are obviously zero by symmetry.

The higher even ones (e.g.\ the fourth, sixth etc.) can be evaluated through brute force calculation or through the application of Wick's theorem. Simply connect up all pairs of $x$s and write down the covariance matrix for each pair. Example:
\begin{eqnarray}
\left<x_{i}x_{j}x_{k}x_{l}\right> & = & \nonumber \left<x_{i}x_{j}\right>\left<x_{k}x_{l}\right>+\left<x_{i}x_{k}\right>\left<x_{j}x_{l}\right>+\left<x_{i}x_{l}\right>\left<x_{j}x_{k}\right> \\
& = & C_{ij}C_{kl} + C_{ik}C_{jl} +C_{il}C_{jk} .
\end{eqnarray}

The number of terms generated in this fashion for the $n$-th order correlation function is $\prod_{i=1}^{n/2}(2i-1)$.

\subsubsection{Marginals and conditionals of Gaussian random fields}

-- Easy computation of marginal and conditional pdfs is a very convenient property of GRFs. First of all, all marginal and conditional densities of GRFs are Gaussian. So all we need to calculate are their means and covariances. Let us the split the GRF up into two parts $x$ and $y$, so that
\begin{equation}
\mu=
\begin{pmatrix}
\mu_{x} \\
\mu_{y}
\end{pmatrix} \quad \mathrm{and} \quad C=
\begin{pmatrix}C_{xx}&C_{xy}\\ C_{yx}&C_{yy}
\end{pmatrix}.
\end{equation}

$C_{xy}=C_{yx}$ by symmetry.

First for the marginal pdfs,
\begin{eqnarray}
\mu_{x} & = & \mu_{x}, \\
C_{xx} & = & C_{xx}, \\
\mu_{y} & = & \mu_{y}, \\
C_{yy} & = & C_{yy}.
\end{eqnarray}

We know these expressions are entirely tautological, but the point is obvious: the marginal mean and marginal covariances are just the corresponding parts of the joint mean and covariance. 

Less trivially, here are the parameters of the conditional densities:
\begin{eqnarray}
\mu_{x|y} & = & \mu_x+	C_{xy}C^{-1}_{yy}(y-\mu_{y}),\\
C_{x|y} & = & C_{xx}-C_{xy}C^{-1}_{yy}C_{yx},\\
\mu_{y|x} & = & \mu_y+	C_{yx}C^{-1}_{xx}(x-\mu_{x}),\label{eq:conditional_mean}\\
C_{y|x} & = & C_{yy}-C_{yx}C^{-1}_{xx}C_{xy}.\label{eq:conditional_covariance}
\end{eqnarray}

From these expressions it is easy to see that for GRFs, lack of covariance implies \textit{independence}, i.e. $p(x,y)=p(x)p(y)$. This is most certainly not the case for general random fields.

These simple formulae are the basis of all forms of optimal filtering (in the least square sense) and the entire Bayesian linear model (see $\S$ \ref{sec:The linear physics CMB time-machine} for an example). They are also behind most of the ideas in scientific data compression, interpolation, extrapolation, and many surprisingly powerful data analysis tools.

\subsubsection{(Mis)-conceptions about Gaussian random fields}

-- In summary, GRFs have nice mathematical properties: they are completely specified by their mean and covariance, all moments exist and can be easily calculated, as well as marginals and conditionals, using only linear algebra. In addition, GRFs are ubiquitous in physics because of the central limit theorem: sums of independent random variates, even non-Gaussian ones, tend to be Gaussian.

This simplicity about GRFs means that it is very easy to work with them, but might also be responsible for some misconceptions. The following lists a set of common and confusing misstatements about (or even definitions of) GRFs in the literature, which often arise from conflating homogeneity and isotropy with Gaussianity.

\begin{description}
\item[Gaussian fields have independent Fourier (or momentum) modes:] 
No: the authors confuse homogeneity and Gaussianity. The statement is false in general, it is only true for homogeneous random fields.
\item[Gaussian random fields are defined as fields with ``random phases":] 
No. First of all, the statement is imprecise and poorly defined (``random?"). The phases that are being referred to are the angles of the complex Fourier amplitudes. The correct argument works as follows. \textit{If} a field is homogeneous \textit{and} Gaussian, \textit{then} the real and imaginary parts are independent, and the phases are independently drawn from a uniform distribution on $[0,2\pi]$. This is something that \textit{follows} for a Gaussian field. However, the argument does not work in reverse. It is trivial to construct examples with independent (between different $k$), uniformly distributed phases in $[0,2\pi]$, where the real and imaginary parts are not independent and are drawn to give a non-Gaussian field.
\item[Histograms of Gaussian Random Fields are Gaussian:]
This is not true in general. A simple counter-example can be constructed with a collection of independent Gaussians with different variances. The histogram of a Gaussian random field with different variances in different pixels will be a sample from a \textit{mixture} of Gaussians with the marginal means and variances, 
\begin{equation}
\mathrm{hist} \leftarrow \frac{1}{n}\sum_{i=1}^{n}g(x_{i}|\mu_{i},C_{ii}).
\end{equation}
So, depending on the GRF, one can obtain any pdf that can be represented in this form. Note that this is only Gaussian if all the marginal means and variances are the same, but \textit{not} in any other case.
\end{description}

These misconceptions show that Gaussianity must be well understood to avoid confusion. In particular, caution is necessary with \textit{tests} of Gaussianity (see $\S$ \ref{sec:Planck results on primordial non-Gaussianity}), as they can easily be confused with tests of inhomogeneity or anisotropy.

\subsection{The linear physics CMB time-machine}
\label{sec:The linear physics CMB time-machine}

In this paragraph, we exemplify the notions reviewed on Gaussian random fields in $\S$ \ref{sec:Gaussian random fields} in a cosmological context: we use the properties of GRFs to build a linear physics ``time-machine" that takes us from the recombination epoch (about 380,000 years after the Big Bang) to the end of inflation ($10^{-35}$ seconds after the Big Bang).

As noted in $\S$ \ref{sec:The birth of perturbations}, calculations of quantum fluctuations produced during the inflationary era predict very nearly Gaussian initial conditions, and -- to very high accuracy -- the CMB is a linear map of the primordial perturbations. Since linear maps preserve Gaussianity, the CMB is a Gaussian field on the sphere. Using this consideration, Komatsu, Spergel and Wandelt (2005) \cite{Komatsu2005} derived the optimal reconstruction for the primordial perturbations on any spherical slice from the cosmic microwave background anisotropy in the limit of small non-Gaussianity (see \cite{Yadav2005} for the extension to polarization).

Let us call $d$ the GRF consisting of the temperature contrast in our CMB sky (our data) and $\Phi$ the gravitational potential (i.e. the primordial curvature perturbation, up to a coefficient). We can see the joint field $\begin{pmatrix}
d \\
\Phi
\end{pmatrix}$ as a GRF with zero mean and covariance 
\begin{equation}
C=
\begin{pmatrix}S+N&X\\ X&P_{\Phi\Phi}
\end{pmatrix}.
\end{equation}

In the previous equation, $S$ is the covariance matrix for the signal and $N$ the covariance matrix for the noise, both in our data, $X$ is the cross-correlation between $d$ and $\Phi$, and $P_{\Phi\Phi}$ is the covariance matrix (auto-correlation) for $\Phi$. We assume a model for $S$, $N$ and $X$. We can then use the formulae for the conditional density of $y$ given $x$, eq. \eqref{eq:conditional_mean} and \eqref{eq:conditional_covariance}, to build the optimal inference of $\Phi$ given $d$:
\begin{eqnarray}
\mu_{\Phi|d} & = & X(S+N)^{-1}d, \label{eq:conditional_varphi_mean}\\
C_{\Phi|d} & = & P_{\Phi\Phi} - X(S+N)^{-1}X .\label{eq:conditional_varphi_covariance}
\end{eqnarray}

The optimal estimate for $\Phi$ given $d$ is the conditional mean of $p(\Phi|d)$, eq. \eqref{eq:conditional_varphi_mean}, with the width (or covariance) of this distribution, eq. \eqref{eq:conditional_varphi_covariance}, describing the uncertainty of this estimate. This technique for inferring a GRF from another GRF, based on their cross-correlations, is called Wiener filtering. Note that we know how to build optimal filters for inverting this linear physics problem, because we understand the conditional densities of GRFs.

\begin{figure}
\begin{center}
\small CMB map, T and E combined \quad \quad \quad \quad \quad \quad \quad Primordial curvature perturbations \normalsize
\includegraphics[width=\textwidth]{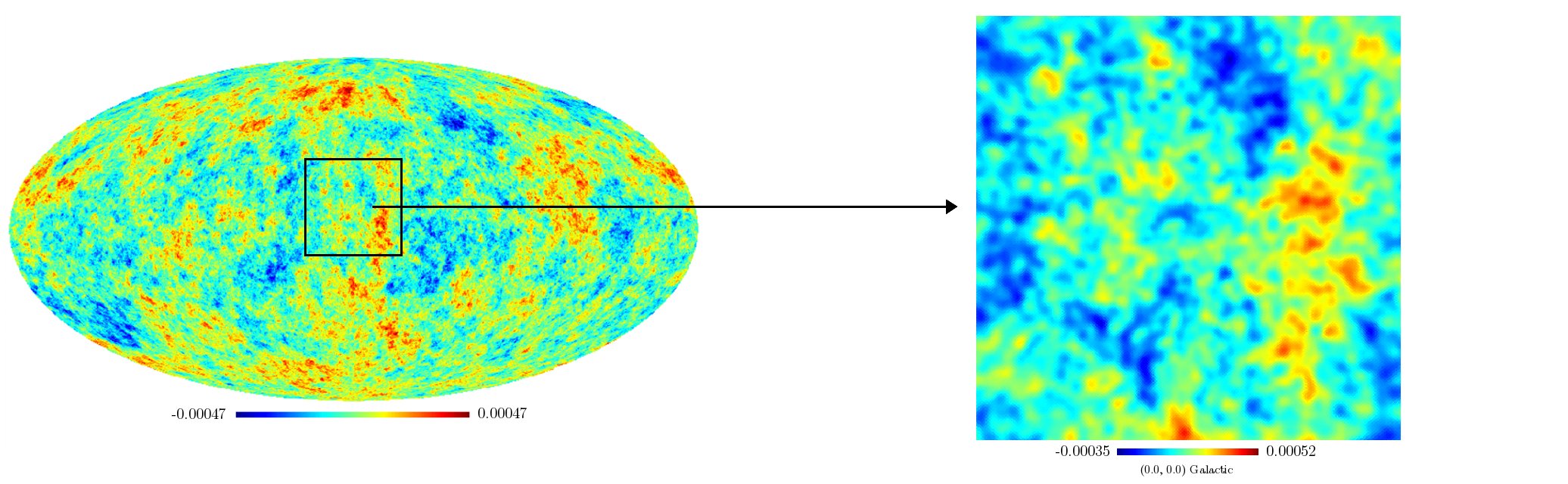}
\end{center}
\caption{Reconstruction of the primordial perturbations from the CMB. On the left, the CMB anisotropies (combination of T and E modes), mapping the Universe about 380,000 years after the Big Bang, and on the right, the reconstructed primordial curvature perturbations (at radius $r=r_{\mathrm{ls}}-160~\mathrm{Mpc}$), just at the end of inflation. Figure adapted from \cite{Yadav2005,Yadav2007}.\label{fig:CMB_time_machine}}
\end{figure}

This approach allow us to infer initial curvature perturbations and to test model predictions for the primordial power spectrum and beyond. Figure \ref{fig:CMB_time_machine} illustrates the reconstruction of primordial curvature fluctuations in some patch of the sky mapped by the CMB anisotropies. The ``CMB time-machine" is an example of the interplay between theory and data: the model predictions we have -- theory -- are tested through data, that in turn give rise to new predictions.

We now understand the reconstruction of the initial conditions in the linear and Gaussian case, but this is just an approximation. Indeed, even for Gaussian fields, non-Gaussian statistics arise for covariance estimation (power spectrum inference, parameter inference). In addition, $21^{\mathrm{st}}$ century cosmology is in an age of \textit{precision} and deals with intrinsically non-linear problems: gravitational non-linearity cannot be ignored at small scales for the analysis of lensing and galaxy surveys, and many theoretical models for inflation involve primordial non-Gaussianity. Non-linearities and non-Gaussianity can also arise in interesting ways when dealing with data imperfection and systematics.

Can we now generalize our inference to the non-linear problem? This is the subject of the following section.
\section{Bayesian cosmostatistics}
\label{sec:Bayesian cosmostatistics}

Statistical reconstruction can be generalized to non-linear settings by dropping the Gaussianity assumption. In the absence of any particular framework, the general idea to go from forward modelling to the inverse problem is that, if one knows how $x$ arises from $y$, then one can use $x$ to constrain $y$:

\begin{equation}
p(y|x)p(x) = p(x|y)p(y) .
\end{equation}

This observation forms the basis of Bayesian statistics.

\subsection{What is Bayesian analysis?}
\label{sec:What is Bayesian analysis?}

Bayesian analysis is a general method for updating the probability estimate for a theory as additional data are acquired. It is based on Bayes' theorem,

\begin{equation}
\label{eq:Bayes}
p(\theta	|d) = \frac{p(d|\theta)p(\theta)}{p(d)} .
\end{equation}

In the previous formula, $\theta$ represents the set of parameters for a particular theory and $d$ the data (before it is known). Therefore, 

\begin{itemize}
\item $p(d|\theta)$ is the probability of the data \textit{before they are known}, given the theory. It is called the \textit{likelihood},
\item $p(\theta)$ is the probability of the theory in the absence of data. It is called the prior probability distribution function or simply the \textit{prior},
\item $p(\theta|d)$ is the probability of the theory	after the data are known. It is called the posterior probability distribution function or simply the \textit{posterior},
\item $p(d)$ is the probability of the data \textit{before they are known}, without any assumption about the theory. It is called the \textit{evidence}.
\end{itemize}

A simple way to summarize Bayesian analysis can be formulated by the following:

\begin{center}
\textit{Whatever is uncertain gets a pdf.}
\end{center}

This statement can be a little disturbing at first (e.g. the value of $\Omega_\mathrm{m}$ is a constant of nature, certainly not a random number of your experiment). What it means is that in Bayesian statistics, pdfs are used to quantify uncertainty of all kinds, not just what is usually referred to as ``randomness'' in the outcome of an experiment. In other words, the pdf for an uncertain parameter can be thought as a ``belief distribution function", quantifying the degree of truth that one attributes to the possible values for some parameter. Certainty can be represented by a Dirac distribution, e.g. if the data determine the parameters completely.

The inputs of a Bayesian analysis are of two sorts:

\begin{itemize}
\item the \textit{prior}: it includes modeling assumptions, both theoretical and experimental. Specifying a prior is a systematic way of quantifying what one assumes true about a theory before looking at the data.
\item the \textit{data}: in cosmology, these can include the temperature in pixels of a CMB map, galaxy redshifts, photometric redshifts pdfs, etc. Details of the survey specifications have also to be accounted for at this point: noise, mask, survey geometry, selection effects, biases, etc.
\end{itemize}

A key point to understand and keep in mind is that the output of a Bayesian analysis is a pdf, the \textit{posterior density}. Therefore, contrary to frequentist statistics, the output of the analysis is not an estimator for the parameters. The word ``estimator'' has a precise meaning in frequentist statistics: it is a function of the data which returns a number that is meant to be close to the parameter it is designed to estimate; or the left and right ends of a confidence interval, etc. The outcome of a Bayesian analysis is the posterior pdf, a pdf whose values give a quantitative measure of the relative degree of rational belief in different parameter values given the combination of prior information and the data. 

\subsection{Prior choice}
\label{sec:Prior choice}

The prior choice is a key ingredient of Bayesian statistics. It is sometimes considered problematic, since there is no unique prescription for selecting the prior. Here we argue that prior specification is not a limitation of Bayesian statistics and does not undermine objectivity as sometimes misstated.

The general principle is that there can be no inference without assumptions, that there does not exist an ``external truth", but that science is building predictive models in certain axiomatic frameworks. In this regard, stating a prior in Bayesian probability theory becomes a systematic way to quantify one's assumptions and state of knowledge about the problem in question before the data is examined. Bayes's theorem gives an unequivocal procedure to update even different degrees of beliefs. As long as the prior has a support that is non-zero in regions where the likelihood is large (Cromwell's rule), the repeated application of the theorem will converge to a unique posterior distribution (Bernstein-von Mises theorem). Generally, objectivity is assured in Bayesian statistics by the fact that, if the likelihood is more informative than the prior, the posterior converges to a common function.

\begin{figure}
\begin{center}
\includegraphics[width=0.5\textwidth]{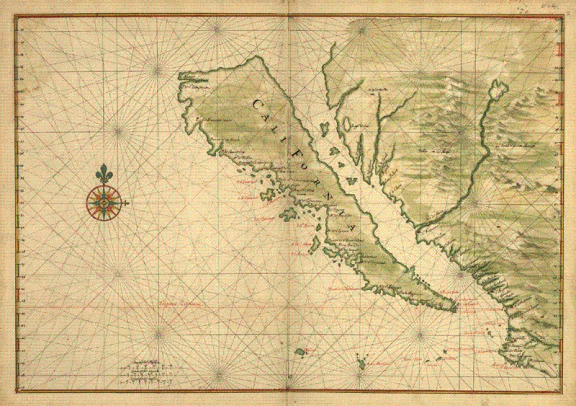}
\end{center}
\caption{What can happen with a bad prior choice, or with a best guess estimator (also in a frequentist framework) instead of the full distribution. The map illustrates a long-held European misconception, that California was not part of mainland North America but rather a large island. This cartographic mistake, one of the most famous in history, dates back to early explorers who misidentified the Baja California peninsula as separated from the continent. A possible explanation is their infusion with the idea that California was a terrestrial paradise, like the Garden of Eden or Atlantis.\label{fig:california}}
\end{figure}

Specifying priors gets assumptions out in the open so that they can be discussed and falsified (see fig. \ref{fig:california}). This is a positive feature of Bayesian probability theory, because frequentists also have to make assumptions that may be more difficult to find within the analysis. An important theorem \cite{Wolpert1997} states that there is ``no free lunch" for optimization problems: when finding the local extremum of a target function (the likelihood in our case) in a finite space, the average performance of any pair of algorithms (that do not resample points) across all possible problems is identical. An important implication is that no universally good algorithm exists; prior information should always be used to match procedures to problems. For this reason, whenever someone (Bayesian or frequentist) tells you ``we did not have to assume anything", do not trust them.

In many situations, external information is highly relevant and should be included in the analysis. For example, when trying to estimate a mass $m$ from some data, one should certainly enforce it to be a positive quantity by setting a prior such that $p(m)~=~0$ for $m~<~0$. Frequentist techniques based on the likelihood can give estimates and confidence intervals that include negative values. Taken at face value, this result is meaningless, unless special care is taken (e.g. the so-called ``constrained likelihood" methods)\footnote{This example illustrates another important point: it is wrong to interpret frequentist confidence intervals as a ``likely range of values the parameter could take'', even though it may only seem natural to demand this kind of information from data analysis. Instead, interpreting frequentist confidence intervals requires imagining a population of repeated experiments in all possible worlds where the parameter takes all possible values. Repeating the analysis for all members of this fictitious population, the $\alpha$-confidence interval will cover the true value of the parameter a fraction $\alpha$ of the time (but there is no sense in which values in the middle are preferred over values at the edge -- in the case of the negative mass example, the true parameter may only ever be in the part of the confidence interval where $m>0$, etc.). }. The use of Bayes's theorem ensures that meaningless results are excluded from the beginning and that you know how to place bets on values of the parameter given the actual data set at hand.

In cosmology, the current state of the art is that previous data (\textsc{COBE}, \textsc{WMAP},~etc.) allowed to establish an extremely solid theoretical footing: the so-called concordance (or $\Lambda$CDM) model. Even when trying to detect deviations from this model in the most recent data (e.g. Planck), it is absolutely well-founded to use it as our prior knowledge about the physical behaviour of the Universe. Therefore, using less informative priors would be refusing to ``climb on the shoulder of giants". In $\S$ \ref{sec:Bayesian non-linear inference of the initial conditions from large-scale structure surveys}, we present an example where combining noisy measurements with a well-motivated physical prior produces a decisive gain in information.

It can happen that the data are not informative enough to override the prior (e.g. for sparsely sampled data or very high-dimensional parameter space), in which case care must be given in assessing how much of the final (first level, see $\S$ \ref{sec:First level inference: Bayesian parameter inference}) inference depends on the prior choice. A good way to perform such a check is to simulate data using the posterior and see if it agrees with the observed data. This can be thought of as ``calculating doubt" \cite{Starkman2008,March2011} to quantify the degree of belief in a model given observational data in the absence of explicit alternative models. Note that even in the case where the inference strongly depends on the prior knowledge, information has been gained on the constraining power (or lack thereof) of the data. In fig. \ref{fig:california} we illustrate an inference based on wrong assumptions.

For model selection questions (second level inference), the impact of the prior choice is much stronger, since it is the available prior volume that matters in determining the penalty that more complex models should incur. Hence, care should be taken in assessing how much of the outcome changes for physically reasonable modifications of the prior.

A vast literature about quantitative prescriptions for prior choice exists (see e.g. \cite{Trotta2008,Heavens2009} for reviews in a cosmological context). An important topic concerns the determination of ``ignorance prior" or ``Jeffreys' priors": a systematic way to quantify a maximum level of uncertainty and to reflect a state of indifference with respect to symmetries of the problem considered. While the ignorance prior is unphysical (nothing is ever completely uncertain) it can be viewed as a convenient approximation to the problem of carefully constructing an accurate representation of weak prior information, which can be very challenging -- especially in high dimensional parameter spaces. 

For example, it can be shown that, if one is wholly uncertain about the position of the pdf, a ``flat prior" should be chosen. In this case, the prior is taken to be constant (within some minimum and maximum value of the parameters so as to be proper, i.e. normalizable to unity). In this fashion, equal probability is assigned to equal states of knowledge. However, note that a flat prior on a parameter $\theta$ does not necessarily correspond to a flat prior on a non-linear function of that parameter, $\varphi(\theta)$. Since $p(\varphi)~=~p(\theta)~\times~|\mathrm{d}\theta/\mathrm{d}\varphi|$, a non-informative (flat) prior on $\theta$ can be strongly informative about $\varphi$. Analogously, if one is entirely uncertain about the width of the pdf, i.e. about the scale of the inferred quantity $\theta$, it can be shown that the appropriate prior is $p(\theta) \propto 1/\theta$, which gives the same probability in logarithmic bins, i.e. the same weight to all orders of magnitude. 

\begin{figure}
\begin{center}
\includegraphics[width=\textwidth]{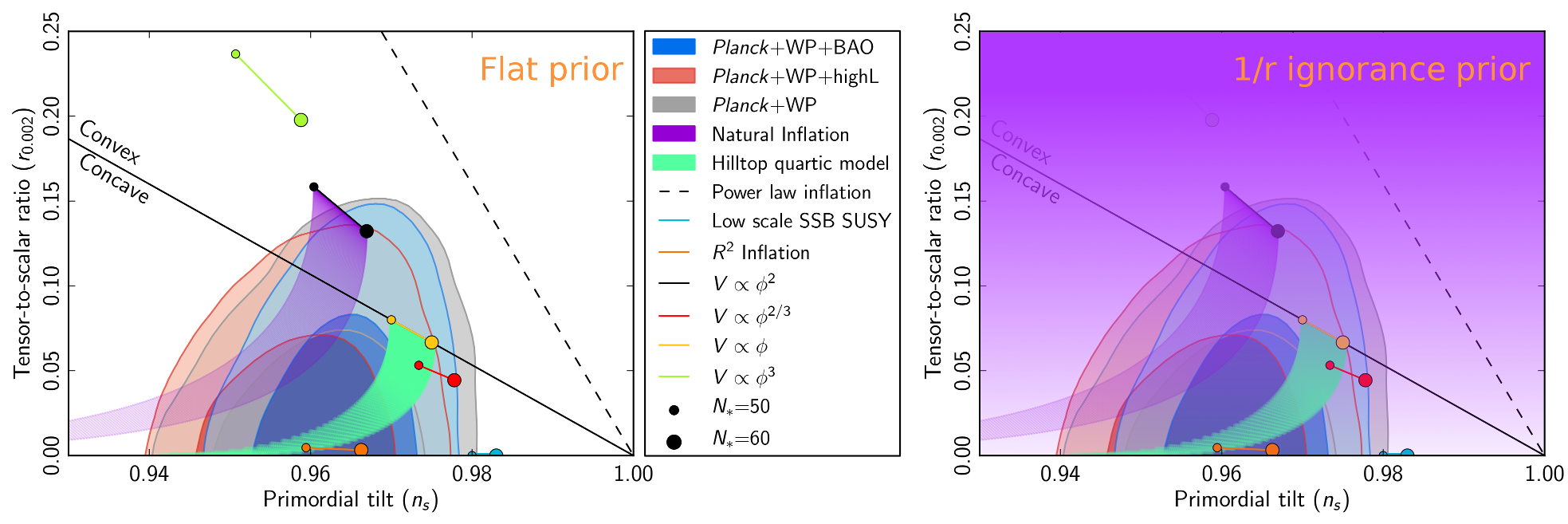}
\end{center}
\caption{Prior effects on inflation constraints from Planck. The plots show the 1-$\sigma$ and 2-$\sigma$ confidence level contours for the tensor-to-scalar ratio $r$ versus the primordial tilt $n_\mathrm{s}$. On the left, the analysis is done with a flat prior on the range considered in parameter space, as can be found in the Planck 2013 paper ``Constraints on inflation" \cite{PlanckCollaboration2013Inflation}. On the right, a $1/r$ ignorance prior is overplotted. Such a prior would favor low tensor-to-scalar ratio.\label{fig:Planck_constraints_inflation}}
\end{figure}

In fig. \ref{fig:Planck_constraints_inflation}, we sketch the effect that a different prior choice could have on the Planck 2013 results for inflation. The data constrain the tensor-to-scalar ratio $r < 0.15$ and the primordial tilt $0.94 < n_\mathrm{s} < 0.98$ at 2-$\sigma$ confidence level. The inflationary potential is constrained to be $V < (1.94 \times 10^{16} \,\mathrm{GeV})^4$. The published analysis is done with a flat prior on the range considered in parameter space, but several other choices are physically possible: the no-boundary wave function $\mathrm{e}^{-V(\phi)}$, Ijjas, Steinhardt \& Loeb's ``$1/$Unlikeliness" \cite{Ijjas2013}, or an ignorance prior about the scale of the gravitational waves background, $1/r$. As shown in fig. \ref{fig:Planck_constraints_inflation}, the latter would favor low tensor-to-scalar ratio.

\subsection{First level inference: Bayesian parameter inference}
\label{sec:First level inference: Bayesian parameter inference}

Parameter inference is the first level of Bayesian probabilistic induction. The problem can be stated as follows. Given a physical model $\mathcal{M}$\footnote{In this section, we make explicit the choice of a model $\mathcal{M}$ by writing it on the right-hand side of the conditioning symbol of all pdfs.}, a set of hypotheses is specified in the form of a vector of parameters, $\theta$ (usually they represent some physically meaningful quantity: $\Omega_{\mathrm{\Lambda}}$, $f_{\mathrm{NL}}$, etc.). Together with the model, priors for each parameter must be specified: $p(\theta|\mathcal{M})$. The next step is to construct the likelihood function for the measurement, with a probabilistic, generative model of the data $p(d|\theta,\mathcal{M})$. The likelihood reflects how the data are obtained: for example, a measurement with Gaussian noise will be represented by a normal distribution.

Once the prior is specified and the data is incorporated in the likelihood function, one mechanically gets the posterior distribution for the parameters, integrating all the information known to date, by plugging into Bayes' theorem (eq. \eqref{eq:Bayes}):

\begin{equation}
\label{eq:First_level_inference}
p(\theta|d,\mathcal{M}) \propto p(d|\theta,\mathcal{M}) p(\theta|\mathcal{M}) .
\end{equation}

Note that the normalizing constant $p(d|\mathcal{M})$ (the Bayesian evidence) is irrelevant for first level inference (but fundamental for second level inference, i.e. model comparison, see $\S$ \ref{sec:Second level inference: Bayesian model comparison}).

Usually, the set of parameters $\theta$ can be divided in some physically interesting quantities $\varphi$ and a set of nuisance parameters $\psi$. The posterior obtained by eq. \eqref{eq:First_level_inference} is the joint posterior for $\theta = (\varphi,\psi)$. The marginal posterior for the parameters of interest is written as (marginalizing over the nuisance parameters)

\begin{equation}
p(\varphi|d,\mathcal{M}) \propto \int p(d|\varphi, \psi, \mathcal{M}) p(\varphi,\psi|\mathcal{M}) \, \mathrm{D}\psi .
\end{equation}

This pdf is the final inference on $\varphi$ from the joint posterior. The following step, to apprehend and exploit this information, is to explore the posterior. It is the subject of the next section.

\subsection{Exploration of the posterior}
\label{sec:Exploration of the posterior}

\begin{figure}
\begin{center}
\includegraphics[width=0.3\textwidth]{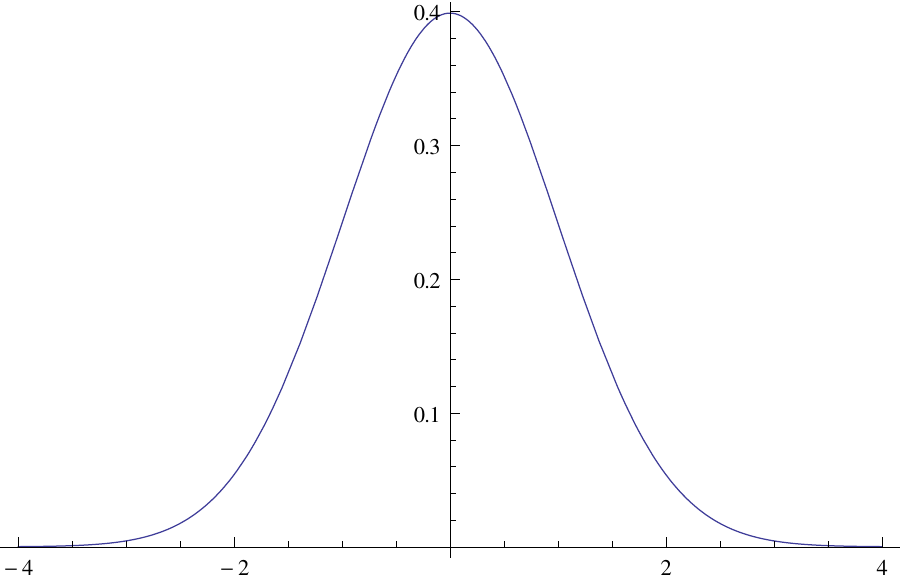} \hspace{0.25cm}\includegraphics[width=0.25\textwidth]{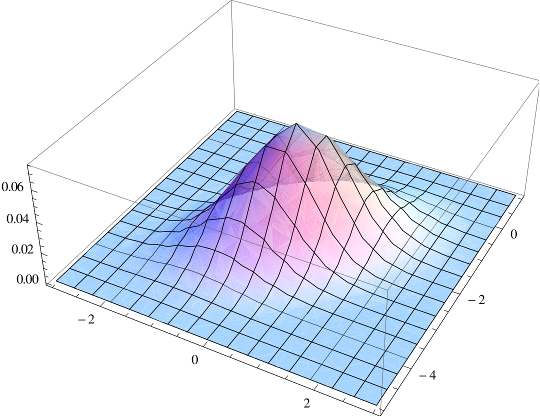}\hspace{0.5cm} \includegraphics[width=0.25\textwidth]{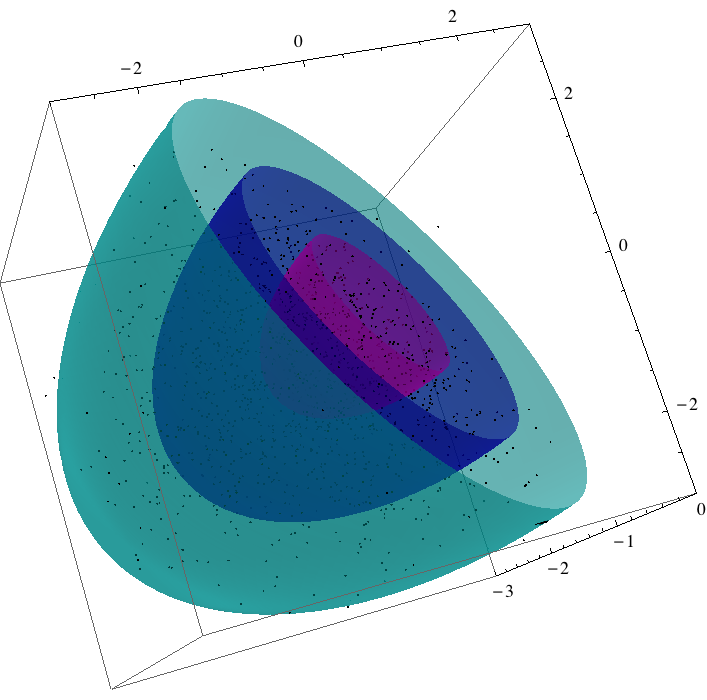}
\end{center}
\caption{Example visualizations of posterior densities in low-dimensional parameter spaces (from left to right: one, two and three).\label{fig:posterior_visualization}}
\end{figure}

The inference of parameters is contained in the posterior pdf, which is the actual output of the statistical analysis. Since this pdf cannot always be easily represented, convenient communication of the posterior information can take different forms:

\begin{itemize}
\item a direct visualization, which is only possible if the parameter space has sufficiently small dimension (see fig. \ref{fig:posterior_visualization}).
\item the computation of statistical summaries of the posterior, e.g. the mean, the median, or the mode of the distribution of each parameter, marginalizing over all others, its standard deviation; the means and covariance matrices of some groups of parameters, etc. It is also common to present the inference by plotting two-dimensional subsets of parameters, with the other components marginalized over (this is especially useful when the posterior is multi-modal or with heavy tails).
\end{itemize}

For typical problems in cosmology, the exploration of a posterior density meets practical challenges, depending on the dimension $D$ of the parameter space. Due to the computational time requirements, numerical evaluation of the posterior density is almost never a smart idea, except for $D<4$. Besides, computing statistical summaries by marginalization means integrating out the other parameters. This is almost never possible analytically (except for Gaussian random fields), and even numerical integration is basically hopeless for $D>5$.

In cosmology, we will often be looking at cases where $D$ is of the order of $10^7$: each pixel value (temperature of the CMB, position or velocity of a galaxy) is a parameter of the analysis. This means that direct evaluation of the posterior is impossible and one has to rely on a numerical approximation: sampling the posterior distribution.

The idea is to approximate the posterior by a set of samples drawn from the real posterior distribution. In this fashion, one replaces the real posterior distribution, $p(\theta|d)$, by the sum of $N$ Dirac delta distributions, $p_N(\theta|d)$:

\begin{equation}
p(\theta|d) \approx p_N(\theta|d) = \frac{1}{N} \sum_{i=1}^{N} \delta_{\mathrm{D}}(\theta - \theta_i) .
\end{equation}

\begin{figure}
\begin{center}
\includegraphics[width=0.6\textwidth]{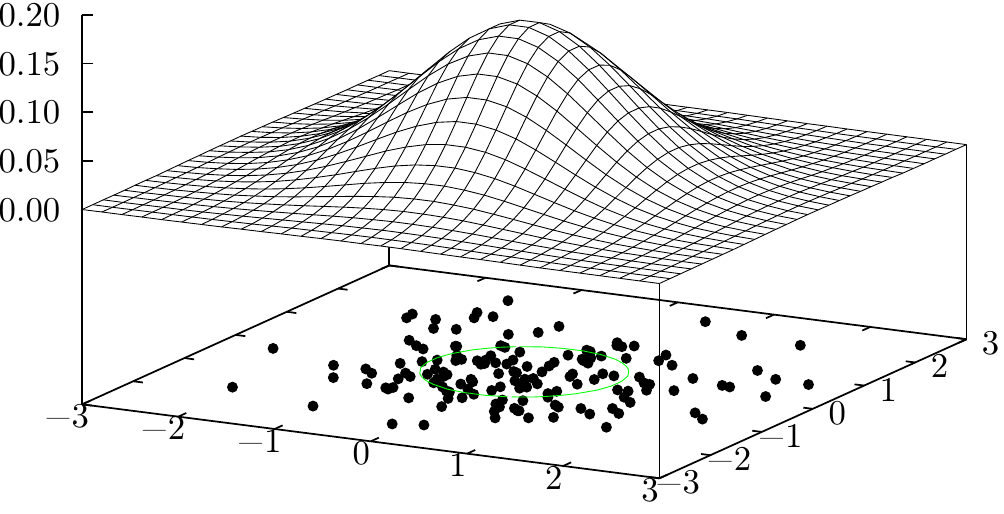}
\end{center}
\caption{Example of a sampled representation of a posterior distribution in two dimensions. A set of samples is constructed in such a way that at any point, the posterior probability is proportional to the local density of samples in parameter space.\label{fig:posterior_sample}}
\end{figure}

A sampled representation of the posterior is constructed in such a way that at any point, the posterior probability is proportional to the local density of samples in parameter space (see fig. \ref{fig:posterior_sample}).

An intuitive way to think about these samples is to consider each of them as a possible version of the truth. The variation between the different samples quantifies the uncertainty that results from having only one Universe (this is a more precise version of the phenomenon known as ``cosmic variance"), incomplete observations (mask, finite volume and number of galaxies, selection effects) and imperfect data (noise, biases, photometric redshifts...). At this point, it is worth stressing that an advantage of Bayesian approach is that it deals with uncertainty independently of its origin, i.e. there is no fundamental distinction between ``statistical uncertainty" coming from the stochastic nature of the experiment and ``systematic uncertainty", deriving from deterministic effects that are only partially known. 

The advantage of a sampling approach is that marginalization over some parameters becomes trivial: one just has to histogram! Specifically, it is sufficient to count the number of samples falling within different bins of some subset of parameters, simply ignoring the values of the others parameters. Integration to get means and variances is also much simpler, since the problem is limited to the computation of discrete sums. More generally, the expectation value of any function of the parameters, $f(\theta)$ is

\begin{equation}
\left\langle f(\theta) \right\rangle \approx \frac{1}{N} \sum_{i=1}^{N} f(\theta_i) .
\end{equation}

Now, how do we get a sampled representation of the posterior? Let us first imagine that we had an infinitely powerful computer. A na\"ive but straightforward sampling algorithm is the following: simulate data from our generative model (draw $\theta$ from the prior, then data from the likelihood knowing $\theta$) and check that the real data agree with the simulated data. If it is the case, keep $\theta$ as one sample, otherwise try again. This is correct in principle, but hugely inefficient. In real life, the standard technique is to use \textit{Markov Chain Monte Carlo}, briefly described in $\S$ \ref{sec:Markov Chain Monte Carlo techniques for parameter inference}.

\subsection{Markov Chain Monte Carlo techniques for parameter inference}
\label{sec:Markov Chain Monte Carlo techniques for parameter inference}

The purpose of Markov Chain Monte Carlo (MCMC) sampling is to construct a sequence of points in parameter space (a so-called ``chain"), whose density is proportional to the posterior density.

A sequence $\left\lbrace X_1, X_2, ..., X_n, ...\right\rbrace $ of random elements of some set (the ``state space") is called a \textit{Markov Chain} if the conditional distribution of $X_{n+1}$ given all the previous elements $X_1$, ... $X_n$ depends only on $X_n$ (the \textit{Markov property}). It is said to have \textit{stationary transition probability} if, additionally, this distribution does not depend on $n$. This is the main kind of Markov chain of interest for MCMC.

Such stationary chains are completely determined by the marginal distribution for the first element $X_1$ (the \textit{initial distribution}) and the conditional distribution of $X_{n+1}$ given $X_n$, called the \textit{transition probability distribution}. 

The crucial property of stationary Markov Chains is that, after some steps depending on the initial position (the so-called ``burn-in" phase), they reach a state where successive elements of the chain are drawn from the high-density regions of the target distribution, in our case the posterior of a Bayesian inference. Exploiting this property, MCMC algorithms use Markovian processes to move from one state to another in parameter space; then, given a set of random samples, they reconstruct the probability heuristically. Several MCMC algorithms exist and the relevant choice is highly dependent on the problem addressed and on the posterior distribution to be explored (see the discussion of the ``no-free lunch" theorem in $\S$ \ref{sec:Prior choice}), but the basic principle is always similar to that of the popular \textsc{CosmoMC} code \cite{Lewis2002}: perform a random walk in parameter space, constrained by the posterior probability distribution. Interestingly, sampling algorithms exist that do not evaluate the posterior pdf (except perhaps occasionally, to maintain high numerical precision). This is particularly useful in high dimensions where it can become prohibitively expensive to evaluate the posterior pdf.

A popular version of MCMC is called the Metropolis-Hastings (MH) algorithm, which works as follows. Let us call $p(\theta)$ the fixed pdf that we want to sample. Initially, one has to choose an arbitrary point $\theta_0$ to be the first sample, and to specify a distribution $q(\theta'|\theta)$ which suggests a candidate $\theta'$ for the next sample value, given the previous sample value $\theta$ ($q$ is called the proposal density or jumping distribution). At each step, one draws a realization $\theta'$ from $q(\theta'|\theta)$ and calculates the acceptance ratio:

\begin{equation}
\label{eq:MH_acceptance}
a=\frac{p(\theta')}{p(\theta)} \frac{q(\theta|\theta')}{q(\theta'|\theta)} .
\end{equation}

\begin{figure}
\begin{center}
\includegraphics[width=0.32\textwidth]{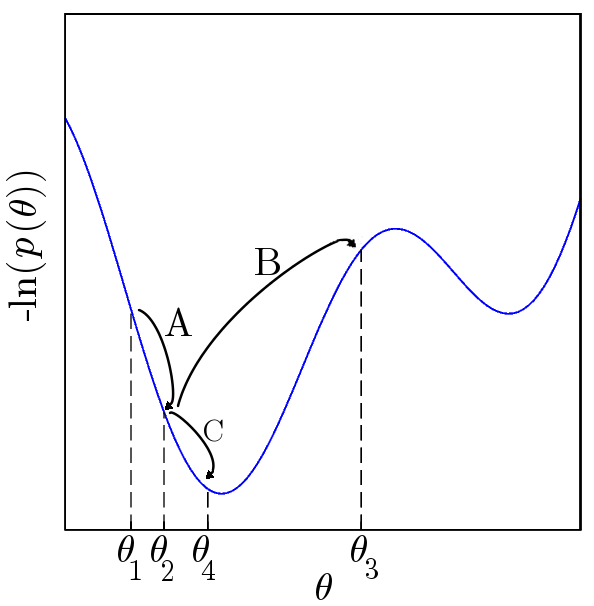} \includegraphics[width=0.32\textwidth]{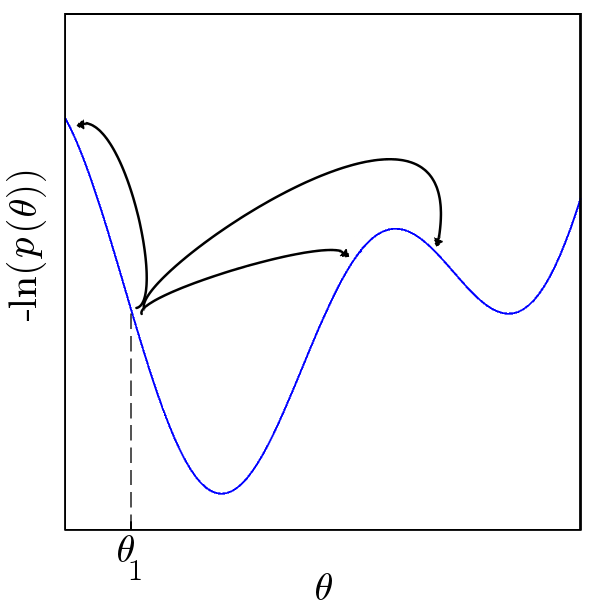} \includegraphics[width=0.32\textwidth]{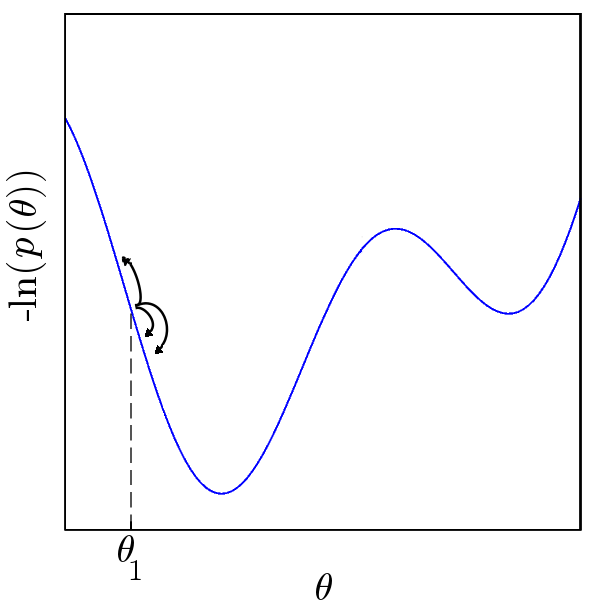}
\end{center}
\caption{\textit{Left panel}. An example of Markov chain constructed by the Metropolis-Hastings algorithm: starting at $\theta_1$, $\theta_2$ is proposed and accepted (step A), $\theta_3$ is proposed and refused (step B), $\theta_4$ is proposed and accepted (step C). The resulting chain is $\left\lbrace \theta_1, \theta_2, \theta_2, \theta_4, ...\right\rbrace$. \textit{Central panel}. An example of what happens with too broad a jump size: the chain lacks mobility because all the proposals are unlikely. \textit{Right panel}. An example of what happens with too narrow a jump size: the chain samples the parameter space very slowly.\label{fig:MH_visualization}}
\end{figure}

If $a>1$, then $\theta'$ is accepted, otherwise it is accepted with probability $a$. In case it is accepted, $\theta'$ becomes the new state of the chain, otherwise the chain stays at $\theta$. A graphical illustration of the MH algorithm is shown in fig. \ref{fig:MH_visualization}. Note that each step only depends on the previous one and is also independent of the number of previous steps, therefore the ensemble of samples of the target distribution, constructed by the algorithm, is indeed a stationary Markov chain.

\begin{figure}
\begin{center}
\begin{tabular}{ccc}
suitable step size & step size too large & step size too small \\
\includegraphics[width=0.32\textwidth]{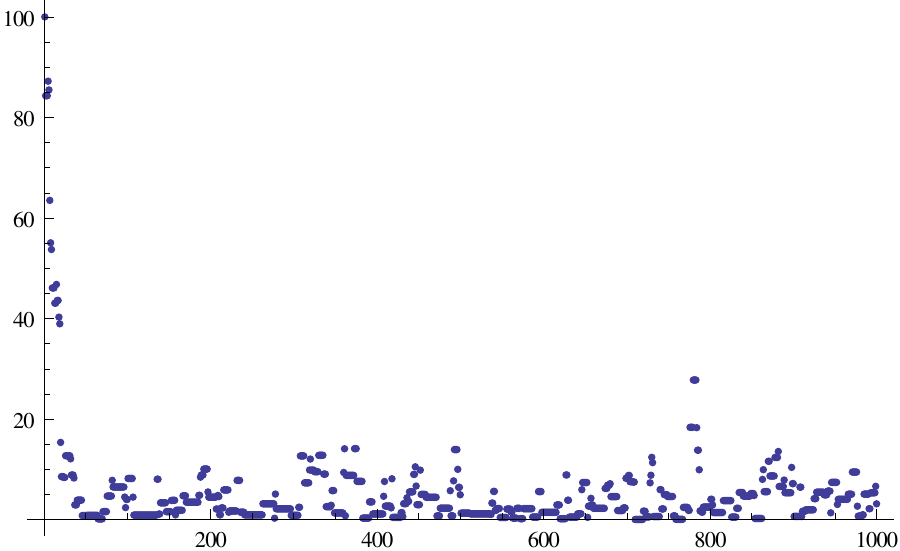} & \includegraphics[width=0.32\textwidth]{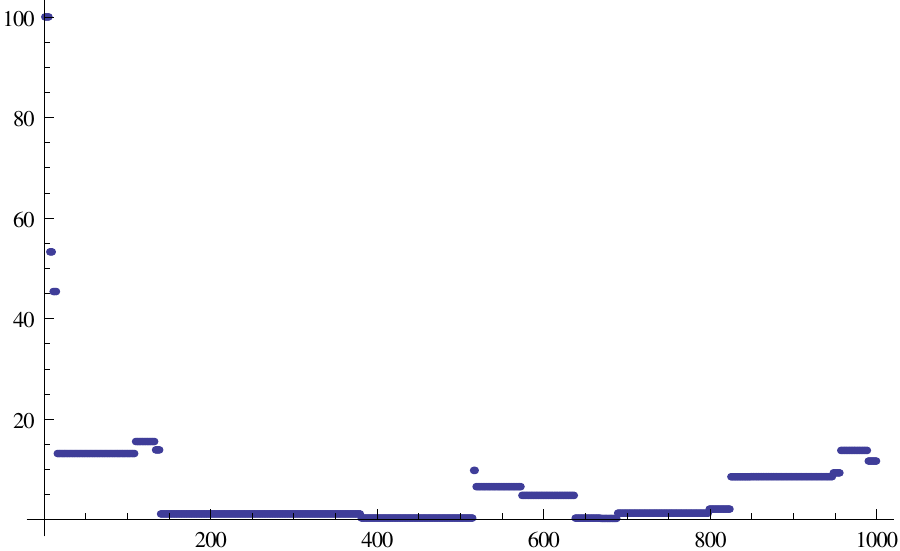} & \includegraphics[width=0.32\textwidth]{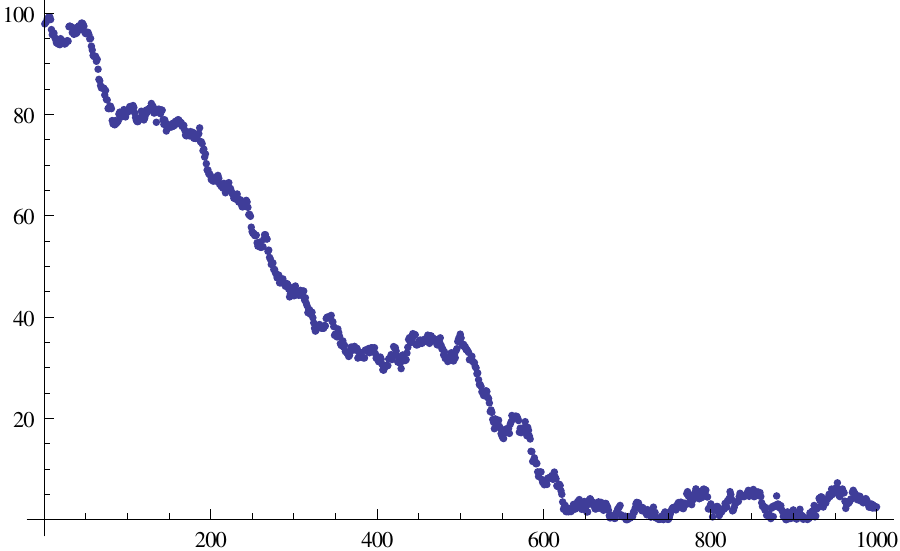}
\end{tabular}
\end{center}
\caption{Example of Markov chains constructed by the Metropolis-Hastings algorithm, sampling the same target distribution but with varying proposal distribution (step size). The ideal behavior with a suitable step size is shown in the left panel. On the central panel, the step size is too large: the maximum likelihood region is not well sampled. On the right panel, the step size is too small: the burn-in phase is very long and the sampling is slow. Note that this phenomena are easily diagnosed using the auto-correlation function of the chain, eq. \eqref{eq:auto-correlation-MC}. \label{fig:MH_visualization_2}}
\end{figure}

In many cases, the MH algorithm will be inefficient if the proposal distribution is sub-optimal. It is often hard to find good proposal distribution if the parameter space has high dimension (e.g. larger than 10). Typically, the chain moves very slowly, either due to a tiny step size, either because only a tiny fraction of proposals are accepted. The initial burn-in phase can be very long, i.e. the chain takes some time to reach high likelihood regions, where the initial position chosen has no influence on the statistics of the chain. Even in the stationary state, sufficient sampling of the likelihood surface can take a very large number of steps. In the central and left panels of fig. \ref{fig:MH_visualization_2}, we illustrate what happens with too broad a jump size (the chain lacks mobility and all proposals are unlikely) or too narrow (the chain moves slowly to sample all the parameter space). Note that the step-size issues can be diagnosed using the lagged auto-correlation function of the chain,

\begin{equation}
\label{eq:auto-correlation-MC}
\xi(\Delta) = \int \theta(t) \theta(t+\Delta) \, \mathrm{d}t .
\end{equation}

A convergence criterion using different chains or sections of chain is proposed in \cite{Gelman1992}. Possible solutions to the issues mentioned involve an adaptive step size or refinements of the standard Metropolis-Hastings procedure. For example, Gibbs sampling is a particular case of the MH algorithm, used when the joint probability distribution is difficult to sample from directly, but the conditional distribution of some parameters given the others is known. It samples an instance from the distribution of each variable in turn, conditional on the current values of the other variables (see e.g. \cite{Wandelt2004} for a cosmological example).

A very efficient MCMC algorithm for high-dimensional problems such as those encountered in cosmology is Hamiltonian Monte Carlo (HMC, originally introduced under the name of hybrid Monte Carlo \cite{Duane1987}). A good general reference is \cite{Neal2011}. HMC interprets the negative logarithm of the posterior as a physical potential, $\psi(\theta) = -\ln(p(\theta))$ and introduces auxiliary variables: ``conjugate momenta" $p_i$ for all the different parameters. Using these new variables as nuisance parameters, one can formulate a Hamiltonian describing the dynamics in the multi-dimensional phase space. Such a Hamiltonian is given as: 
\begin{equation}
\label{eq:HMC_hamiltonian}
H(\left\lbrace \theta_i , p_i \right\rbrace) = \sum_{i,j} \frac{1}{2} p_i M_{ij}^{-1} p_j + \psi(\theta) = - \ln(p(\left\lbrace \theta_i , p_i \right\rbrace)) ,
\end{equation}
where $M$ is a symmetric positive definite ``mass matrix" whose choice can strongly impact the performance of the sampler.

The proposal step for HMC works as follows. One draws a realization of the momenta from the distribution defined by the kinetic energy term, i.e. a multi-dimensional Gaussian with a covariance matrix $M$, then moves $\theta$ using a Hamiltonian integrator in parameter space, respecting symplectic symmetry. In other words, we first ``kick the system" then follow its deterministic dynamical evolution in phase space according to the Hamilton equations, well-known from analytical mechanics.

The acceptance probability for the new point ($\left\lbrace \theta'_i , p'_i \right\rbrace$) follows the usual rule (eq. \eqref{eq:MH_acceptance}, with a symmetric proposal, i.e. $q(\theta') = q(\theta)$):

\begin{equation}
p_a = \min\left[ 1 \, ; \frac{p(\left\lbrace \theta'_i , p'_i \right\rbrace)}{p(\left\lbrace \theta_i, p_i \right\rbrace))} \right] = \min \left[ 1 \, ; \exp(-(H(\left\lbrace \theta'_i , p'_i \right\rbrace) - H(\left\lbrace \theta_i , p_i \right\rbrace))) \right] .
\end{equation}

Since the energy (the Hamiltonian given in eq. \eqref{eq:HMC_hamiltonian}) is conserved, this procedure always provides an acceptance rate of unity! In practice, numerical errors can lead to a somewhat lower acceptance rate but HMC remains computationally much cheaper than standard MH techniques in which proposals are often refused. In the end, we discard the momenta and yield the target parameters by marginalization:

\begin{equation}
p(\left\lbrace \theta_i \right\rbrace) = \int p(\left\lbrace \theta_i, p_i \right\rbrace)) \, \mathrm{D}\left\lbrace p_i \right\rbrace .
\end{equation}

Applications of HMC in cosmology include: the determination of cosmological parameters (in combination with \textsc{Pico}) \cite{Hajian2007}, CMB power spectrum inference \cite{Taylor2008} and Bayesian approach to non-Gaussianity analysis \cite{Elsner2010a}, log-normal density reconstruction \cite{Jasche2010} (including from photometric redshift surveys \cite{Jasche2012}), dynamical, non-linear reconstruction of the initial conditions from galaxy surveys \cite{Jasche2013} (this application is also the first example mentioned in the following section, see $\S$ \ref{sec:Bayesian non-linear inference of the initial conditions from large-scale structure surveys}), joint power spectrum and bias model inference \cite{Jasche2013a}.

\subsection{Second level inference: Bayesian model comparison}
\label{sec:Second level inference: Bayesian model comparison}

In the case where there are several competing theoretical models, second level inference (or Bayesian model comparison) provides a systematic way to estimate their relative probability given the data and any prior information available. It does not replace parameter inference, but rather extends the assessment of hypotheses to the space of theoretical models.

This allow to quantitatively address everyday questions in astrophysics -- Did I detect a source, a spectral line, gravitational waves, an exoplanet, a microlensing event? -- and in cosmology -- Is the Universe flat or should one allow a non-zero curvature parameter? Are the primordial perturbations Gaussian or non-Gaussian? Are there isocurvature modes? Are the perturbations strictly scale-invariant ($n_\mathrm{s} = 1$) or should the spectrum be allowed to deviate from scale-invariance? Is there evidence for a deviation from general relativity? Is the equation of state of dark energy equal to $-1$? 

In many of the situations above, Bayesian model comparison offers a way of balancing complexity and goodness of fit: it is obvious that a model with more free parameters will always fit the data better, but it should also be ``penalized" for being more complex and hence, less predictive. The notion of predictiveness really is central to Bayesian model comparison in a very specific way: the evidence is actually the prior predictive pdf, the pdf over all data sets predicted for the experiment before data are taken. Since predictiveness is a criterion for good science everyone can agree on, it is only natural to compare models based on how well they predicted the data set before it was obtained. This criterion arises automatically in the Bayesian framework.

You may be familiar with the scientific guiding principle known as Occam's razor: the simplest model compatible with the available information ought to be preferred. We now understand this principle as a consequence of using predictiveness as the criterion. A model that is so vague (e.g. has so many parameters) that it can predict a large range of possible outcomes will predict any data set with smaller probability than a model that is highly specific and therefore has to commit to predicting only a small range of possible data sets. It is clear that the specific model should be preferred if the data falls within the narrow range of its prediction. Conversely, we default to the broader more general model only if the data are incompatible with the specific model. Therefore, Bayesian model comparison offers formal statistical grounds for selecting models based on an evaluation whether the data truly favor the extra complexity of one model compared to another. 

Contrary to frequentists goodness-of-fit tests, second level inference always requires an alternative explanation for comparison (finding that the data are unlikely within a theory does not mean that the theory itself is improbable, unless compared with an alternative). The prior specification is crucial for model selection issues: since it is the range of values that parameters can take that controls the sharpness of Occam's razor, the prior should exactly reflect the available parameter space under the model before obtaining the data.

The evaluation of model $\mathcal{M}$'s performance given the data is quantified by $p(\mathcal{M}|d)$. Using Bayes' theorem to invert the order of conditioning, we see that it is proportional to the product of the prior probability for the model itself, $p(\mathcal{M})$, and of the Bayesian evidence already encountered in first level inference, $p(d|\mathcal{M})$:

\begin{equation}
p(\mathcal{M}|d) \propto p(\mathcal{M})p(d|\mathcal{M}) .
\end{equation}

Usually, prior probabilities for the models are taken as all equal to $1/N_\mathrm{m}$ if one considers $N_\mathrm{m}$ different models (this choice is said to be \textit{non-committal}). When comparing two competing models denoted by $\mathcal{M}_1$ and $\mathcal{M}_2$, we are interested in the ratio of their posterior probabilities, also called \textit{conditional odds} or \textit{posterior odds}, given by \textit{Bayes's rule}:

\begin{equation}
\mathcal{O}(\mathcal{M}_1:\mathcal{M}_2|d) \equiv \frac{p(\mathcal{M}_1|d)}{p(\mathcal{M}_2|d)} = \frac{p(\mathcal{M}_1)}{p(\mathcal{M}_2)} \frac{p(d|\mathcal{M}_1)}{p(d|\mathcal{M}_2)} .
\end{equation}

This involves the \textit{marginal odds} or \textit{prior odds}, $\mathcal{O}(\mathcal{M}_1:\mathcal{M}_2) \equiv p(\mathcal{M}_1)/p(\mathcal{M}_2)$. With non-committal priors on the models, $p(\mathcal{M}_1) = p(\mathcal{M}_2)$, the ratio simplifies to the ratio of evidences, called the \textit{Bayes factor},

\begin{equation}
\mathcal{B}_{12} \equiv \frac{p(d|\mathcal{M}_1)}{p(d|\mathcal{M}_2)} .
\end{equation}

The Bayes factor is the appropriate quantity to update our relative state of belief in two competing models in light of the data, regardless of the relative prior probabilities we assign to them: a value of $\mathcal{B}_{12}$ greater than one means that the data give a better support in favor of model $\mathcal{M}_1$ versus model $\mathcal{M}_2$. Note that the Bayes factor is very different from the ratio of the likelihoods: a more complicated model will always yield higher likelihood values, whereas the evidence will favor a simpler model if the fit is nearly as good, through the smaller prior volume.

Conditional odds (or directly the Bayes factor in case of non-committal priors) are often interpreted against the Jeffreys's scale for the strength of evidence. For two competing models $\mathcal{M}_1$ and $\mathcal{M}_2$ with non-committal priors ($p(\mathcal{M}_1) = p(\mathcal{M}_2) = 1/2$) and exhausting the model space ($p(\mathcal{M}_1|d) + p(\mathcal{M}_2|d) = 1$), the relevant quantity is the logarithm or the Bayes factor, $\ln \mathcal{B}_{12}$ for which thresholds at values of 1.0, 2.5 and 5.0 are set (corresponding to odds of about 3:1, 12:1 and 150:1, representing weak, moderate and strong evidence, respectively). The use of a logarithm in this empirical scale quantifies the principle that the evidence for a model only accumulates slowly with new informative data: rising up one level in the evidence strength requires about one order of magnitude more support. It is interesting note that, due to its threshold nature, the trustworthiness of the Jeffreys's scale in a cosmological context is still debated \cite{Nesseris2013}.

An important particular case is when $\mathcal{M}_2$ is a simpler model, described by fewer ($n'<n$) parameters than $\mathcal{M}_1$. $\mathcal{M}_2$ is said to be \textit{nested} in model $\mathcal{M}_1$ if the $n'$ parameters of $\mathcal{M}_2$ are also parameters of $\mathcal{M}_1$. $\mathcal{M}_1$ has $p \equiv n-n'$ extra parameters that are fixed to fiducial values in $\mathcal{M}_2$. For simplicity, let us assume that there is only one extra parameter $\zeta$ in model $\mathcal{M}_1$, fixed to 0 in $\mathcal{M}_2$ ($\zeta$ describes the continuous deformation from one model to the other). Let us denote the set of other parameters by $\theta$. Under these hypotheses, the evidence for $\mathcal{M}_1$ is $p(d|\mathcal{M}_1) \equiv p(d|\mathcal{M}_{\theta,\zeta})$ and the evidence for $\mathcal{M}_2$ is $p(d|\mathcal{M}_2)~\equiv~p(d|\mathcal{M}_{\theta,\zeta=0})~=~p(d|\zeta=0, \mathcal{M}_{\theta,\zeta})$. We also assume non-committal priors for $\mathcal{M}_1$ and $\mathcal{M}_2$.

If the prior for the additional parameter $\zeta$ is independent of the other parameters (which makes the joint prior separable: $p(\zeta,\theta|\mathcal{M}_{\theta,\zeta}) = p(\zeta|\mathcal{M}_{\theta,\zeta})p(\theta|\mathcal{M}_{\theta,\zeta=0})$), it can be shown that the Bayes factor takes a simple form, the Savage-Dickey ratio \cite{Dickey1971,Verdinelli1995}
\begin{equation}
\mathcal{B}_{12} = \frac{p(d|\mathcal{M}_{\theta,\zeta})}{p(d|\mathcal{M}_{\theta,\zeta=0})} = \frac{p(\zeta=0|\mathcal{M}_{\theta,\zeta})}{p(\zeta=0|d,\mathcal{M}_{\theta,\zeta})} ,
\end{equation}
that is, the ratio of the marginal prior and the marginal posterior of the larger model $\mathcal{M}_{1}$, where the additional parameter $\zeta$ is held at its fiducial value. The Bayes factor favors the ``larger'' model only if the data decreases the posterior pdf at the fiducial value compared to the prior. Operationally, if $n-n'$ is small, one can easily compute the Savage-Dickey ratio given Monte Carlo samples from the posterior and prior of $\mathcal{M}_{1}$ by simply estimating the marginal densities at the fiducial value.
\section{Applications of inference}
\label{sec:Applications of inference}

In this section we discuss two applications of inference designed to learn about the initial conditions of the Universe, based on large-scale galaxy surveys ($\S$ \ref{sec:Bayesian non-linear inference of the initial conditions from large-scale structure surveys}) and the cosmic microwave background ($\S$ \ref{sec:Planck results on primordial non-Gaussianity}). Between these two approaches, we expect the CMB to have much more signal on very large scales and to be more easily interpreted using linear physics (see $\S$ \ref{sec:The linear physics CMB time-machine}), but in principle, tracers of the density field \textit{should} win overall, simply because there are vastly more perturbation modes in a three-dimensional data set, which greatly reduces sample variance.

\subsection{Bayesian non-linear inference of the initial conditions from large-scale structure surveys}
\label{sec:Bayesian non-linear inference of the initial conditions from large-scale structure surveys}

A natural idea for the application of Bayesian non-linear inference is the reconstruction of the initial conditions from large-scale structure surveys, allowing the construction of a \textit{non-linear} ``time-machine" using posterior exploration. Ideally, the analysis should be formulated in terms of the simultaneous constraints that surveys place on the initial density field and the physical evolution that links the initial density field to the observed tracers of the evolved density field. Due to the complicated relationship between the distribution of tracers and the underlying mass distribution, the current state of the art of statistical analyses of LSS surveys is far from this ideal.

These complications arise from the lack of a detailed model of the way galaxies arise in response to the spatial fluctuations in the dark matter distribution, which involves very intricate physics (the ``bias" problem). In addition, even for dark matter alone, the density field has undergone non-linear dynamical evolution at late times, on scales smaller than $\sim20$ Mpc/$h$, which has coupled the perturbation modes and erased information about the mode amplitudes in the initial conditions. Finally, uncertainties arise from the incompleteness of the observations and the imperfectness of the experiment.

Incorporating a fully non-linear evolution into cosmological inference, let alone a full physical model of galaxy formation, is not computationally tractable. Therefore, the challenge is to produce an analysis with uncontroversial prior information and an approximate physical model, insensitive to the complications described above. This allows a robust, non-linear inference, including the reconstitution of some of the information that has not been captured by the data.

In \cite{Jasche2012}, progress is described towards the exploration of a set of three-dimensional initial conditions that are consistent with the galaxy distribution sampling the final density field, and corresponding dynamical histories (see also \cite{Kitaura2013,Wang2013}). The implementation of the initial conditions sampler is called \textsc{borg} (Bayesian Origin Reconstruction from Galaxies). The parameter space consists of the value in each of the voxels of the 3D initial density field (about $10^7$ parameters). As a refinement, the joint inference can also include the cosmological power spectrum and luminosity dependent galaxy biases \cite{Jasche2013}.

The physical model for the gravitational dynamics connecting the initial conditions with the final density field is second-order Lagrangian perturbation theory (2LPT). This model, valid in the linear and mildly non-linear regimes of cosmic structure formation, has been widely applied in data analysis and for fast generation of galaxy mock catalogs. It provides a good model for the one and two-point functions of the field, but also reproduces reasonably well features that are associated with higher-order correlators, such as walls and filaments \cite{Moutarde1991,Buchert1994,Bouchet1995,Scoccimarro2000,Scoccimarro2002}. Of course, the validity of this approach ceases, once the evolution of the large-scale structure enters the multi-stream regime. Any numerically efficient and flexible extension of 2LPT, pushing dynamic analyses of the large scale structure further into the non-linear regime, permits a significant gain of information \cite{Leclercq2013}.

The prior information for the primordial perturbations is chosen as a Gaussian random field; it is the best observationally-supported physical model for the initial conditions (see $\S$ \ref{sec:The birth of perturbations}). Therefore, the prior for the evolved density is the initial Gaussian density field evolved by a 2LPT model. The data is modeled as a Poisson sample from the evolved density fields (in this approach, galaxies are considered as matter tracers so that the statistical uncertainty due to the discrete nature of their distribution can be modeled as an inhomogeneous Poissonian process), which yields a Poissonian likelihood distribution.
 
The sampler for the posterior distribution uses an efficient implementation of the Hamiltonian Markov Chain Monte Carlo method. It accurately accounts for all non-linearities and non-Gaussianities involved in the inference process and achieves high statistical efficiency, even in low signal to noise regimes. The accept-reject method stays computationally feasible, in spite of the high dimensionality of the problem, due to Hamiltonian dynamics which would yield an acceptance rate of unity in the absence of numerical errors.

\begin{figure}
\begin{center}
\includegraphics[width=\textwidth]{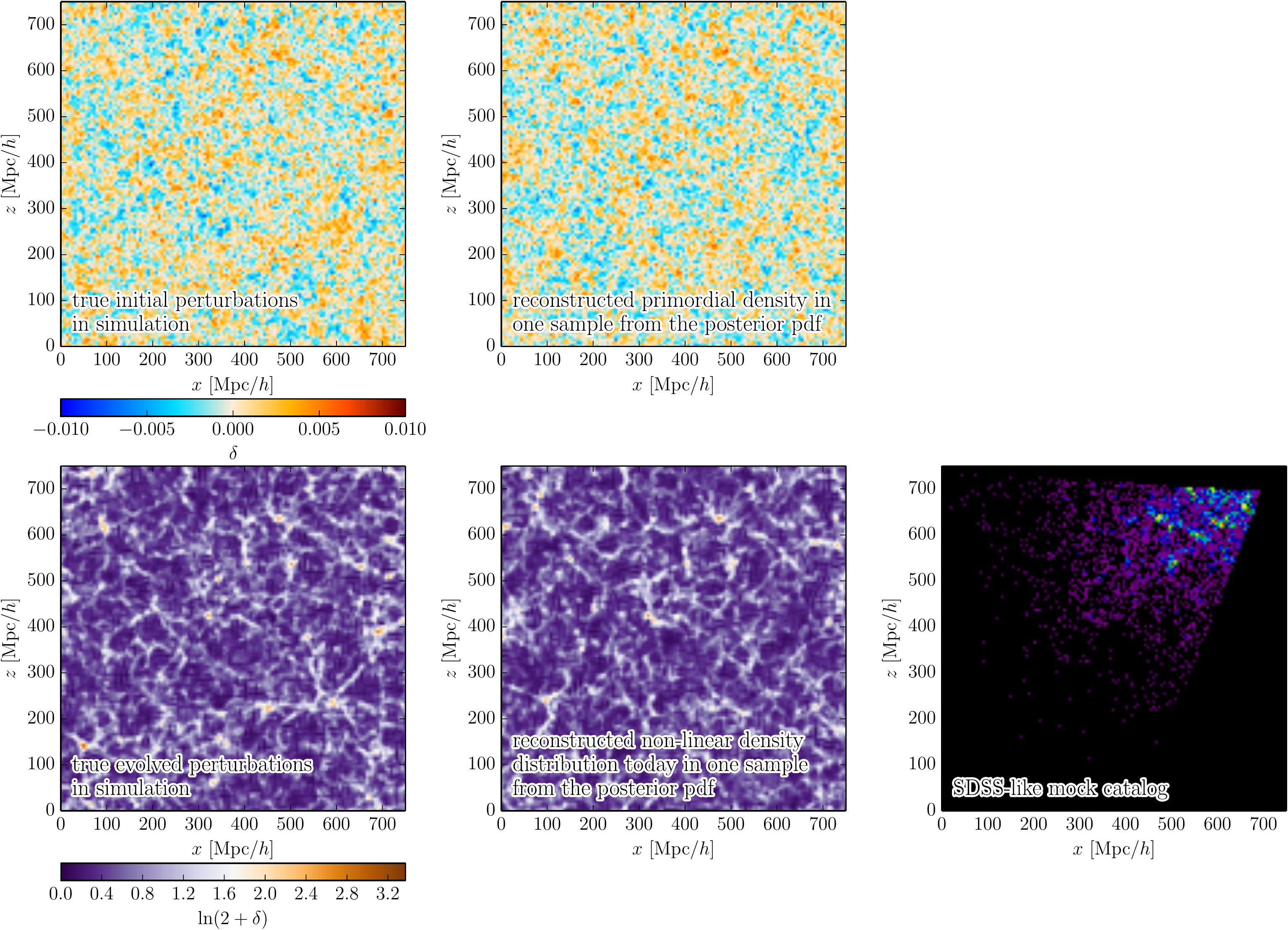} 
\end{center}
\caption{Bayesian physical reconstruction of initial conditions: proof of concept for the \textsc{borg} code. A simulated density field (left panel) is used for the generation of a SDSS-like mock catalog (right panel). In the central panel, the \textsc{borg} reconstruction of the initial (upper figure) and final (lower figure) density, in one sample of the posterior pdf, is shown for comparison with the true perturbations. Note that  structures are well reconstructed, both in the initial and final conditions, in the region constrained by observations (in the upper right part in this slice); and are allowed to fluctuate consistently with the physical model, in the unconstrained zones. Figure adapted from \cite{Jasche2013}, courtesy of Jens Jasche.\label{fig:borg_test}}
\end{figure}

Tests of \textsc{borg} against simulated data show that the physical 2LPT model, though approximate, permits the accurate inference of the present-day density field on scales larger than $\sim 6$ Mpc/$h$ (including non-linear features such as walls and filaments). Reconstructed initial conditions show statistical consistency with the Gaussian simulation inputs. Tests against realistic mock catalogs demonstrate the robustness of the reconstruction, as well as the numerical feasibility to apply this approach to the current generation of available surveys. In fig. \ref{fig:borg_test}, we illustrate the performance of \textsc{borg}, demonstrating the correlation between the final density field and the data as well as the connection between structures in the initial conditions and in the final density field.

This inference problem is of particular interest because it shows how combining millions of noisy measurements of galaxies with a physical prior, namely the Gaussianity of the initial conditions, produces a decisive gain in information.

\subsection{Planck results on primordial non-Gaussianity}
\label{sec:Planck results on primordial non-Gaussianity}

As we have already mentioned, the CMB is at very good accuracy, a GRF on the sphere, to the extent that primordial perturbations are Gaussian, transfer is linear, and late-time effects are subtracted. However, there is much more information in the CMB temperature map than the power spectrum of fluctuations alone. Since the expected three-point correlation features of standard inflation (single-field, slow-roll, with standard kinetic term and initial Bunch-Davies vacuum) are well known \cite{Maldacena2003,Creminelli2004}, non-Gaussianity (NG) is currently the highest precision test of standard inflation.

The Planck satellite probes non-Gaussianity at the level of $0.01\%$. For comparison, flatness is constrained at a level of $\sim 0.1 \%$, and isocurvature modes at a level of $\sim 1 \%$. The detailed analysis can be found in the Planck 2013 papers ``Constraints on primordial non-Gaussianity" \cite{PlanckCollaboration2013NonGaussianity}, ``Constraints on inflation" \cite{PlanckCollaboration2013Inflation} and ``The integrated Sachs-Wolfe effect" \cite{PlanckCollaboration2013ISW} (for the late-time bispectrum).

\subsubsection{Bispectra, the inflationary landscape and non-Gaussianity}

-- In a GRF, all the information is encoded in the two-point correlation function. The lowest-order statistic to probe non-Gaussian features is therefore the three-point correlation function or its Fourier space counterpart, the bispectrum. Using the transfer function (eq. \eqref{eq:transfer_function}), the CMB bispectrum in harmonic space, $B^{\ell_1 \ell_2 \ell_3}_{m_1 m_2 m_3} \equiv \left< a_{\ell_1 m_1} a_{\ell_2 m_2} a_{\ell_3 m_3} \right>$, and the angle-averaged bispectrum,

\begin{equation}
B_{\ell_1 \ell_2 \ell_3} \equiv \sum_{m_1,m_2,m_3} \begin{pmatrix}
   \ell_1 & \ell_2 & \ell_3 \\
   m_1 & m_2 & m_3
  \end{pmatrix} B^{\ell_1 \ell_2 \ell_3}_{m_1 m_2 m_3} ,
\end{equation}

\noindent are directly related to the primordial bispectrum, defined by (assuming homogeneity and isotropy)

\begin{equation}
\left< \Phi(\textbf{k}_1) \Phi(\textbf{k}_2) \Phi(\textbf{k}_3) \right> = (2\pi)^3 \, \delta_\mathrm{D}(\textbf{k}_1+\textbf{k}_2+\textbf{k}_3) \, B_\Phi(k_1,k_2,k_3) .
\end{equation}

Generally, the bispectrum can be written as

\begin{equation}
B_\Phi(k_1,k_2,k_3) = f_\mathrm{NL} \, F(k_1,k_2,k_3), 
\end{equation}

\noindent where the function $F(k_1,k_2,k_3)$ describes the \textit{shape} of the bispectrum, i.e. the dependence on the type of triangle formed by the vectors $\textbf{k}_1, \textbf{k}_2, \textbf{k}_3$, and its \textit{running}, i.e. its dependence on scale. The dimensionless ``non-linearity" parameter $f_\mathrm{NL}$ \cite{Gangui1994} characterizes the amplitude of non-Gaussianity. The bispectrum is usually normalized to the so-called reduced bispectrum $b_{\ell_1 \ell_2 \ell_3}$, defined by (e.g. \cite{Liguori2010})

\begin{equation}
B_{\ell_1 \ell_2 \ell_3} \equiv \sqrt{\frac{(2\ell_1+1)(2\ell_2+1)(2\ell_3+1)}{4\pi}} \begin{pmatrix}
   \ell_1 & \ell_2 & \ell_3 \\
   0 & 0 & 0
  \end{pmatrix} b_{\ell_1 \ell_2 \ell_3} B_{\ell_1 \ell_2 \ell_3}^{m_1 m_2 m_3} .
\end{equation}

Non-Gaussian features of a random field are usually divided into different kinds, according to the shape of triangle for which the bispectrum peaks. A sizable amount of NG with specific triangular configuration is produced if any of the assumptions of standard inflation (single-field, canonical kinetic term, slow-roll, initially lying in a Bunch-Davies vacuum state) is violated:

\begin{itemize}
\item local NG (where the signal peaks for ``squeezed" triangles: $k_1 \ll k_2 \simeq k_3$) are often produced in multi-field models \cite{Bartolo2002,Bernardeau2002}, in particular the curvaton scenario \cite{Linde1997,Lyth2002,Lyth2003}, and in some alternative scenarii to inflation, for instance in cyclic/ekpyrotic models \cite{Khoury2001,Steinhardt2002,Lehners2008}.
\item equilateral NG (where the signal peaks for $k_1 \simeq k_2 \simeq k_3$) are produced in single-field models with a non-canonical kinetic term \cite{Chen2007} (e.g. $k$-inflation \cite{Armendariz-Picon1999,Garriga1999}, Dirac-Born-Infeld -- DBI -- inflation \cite{Silverstein2004, Alishahiha2004}), in models with higher-derivative interactions in the inflationary Lagrangian (e.g. ghost inflation \cite{Arkani-Hamed2004}) and models arising from effective field theories \cite{Cheung2008}.
\item folded NG (where the signal peaks for ``flat" triangles: $k_1~\simeq~k_2~\simeq~k_3/2$) are produced in single-field models initially lying in a non-Bunch-Davies vacuum \cite{Chen2007,Holman2008} and in models with general higher-derivative interactions \cite{Senatore2010,Bartolo2010}
\item orthogonal NG distinguishes between different variants of non-canonical kinetic term \cite{Senatore2010} and higher derivative interactions. It is also a prediction of Galileon inflation \cite{Nicolis2009, Deffayet2009, deRham2010}.
\end{itemize}

\begin{figure}
\begin{center}
\includegraphics[width=0.32\textwidth]{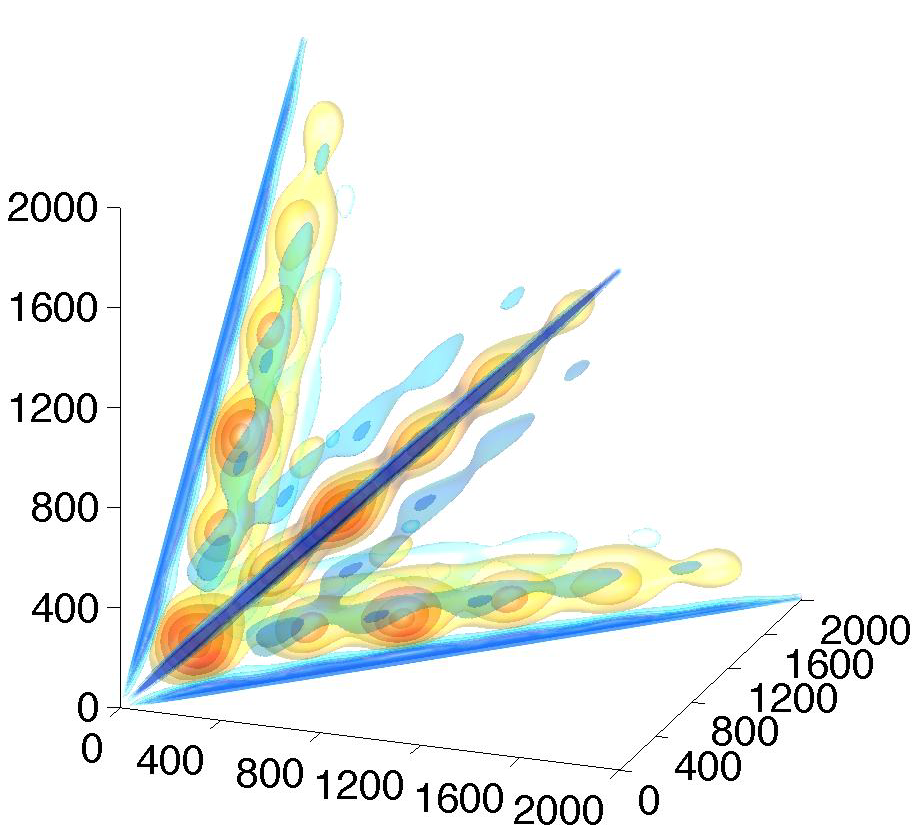} 
\includegraphics[width=0.32\textwidth]{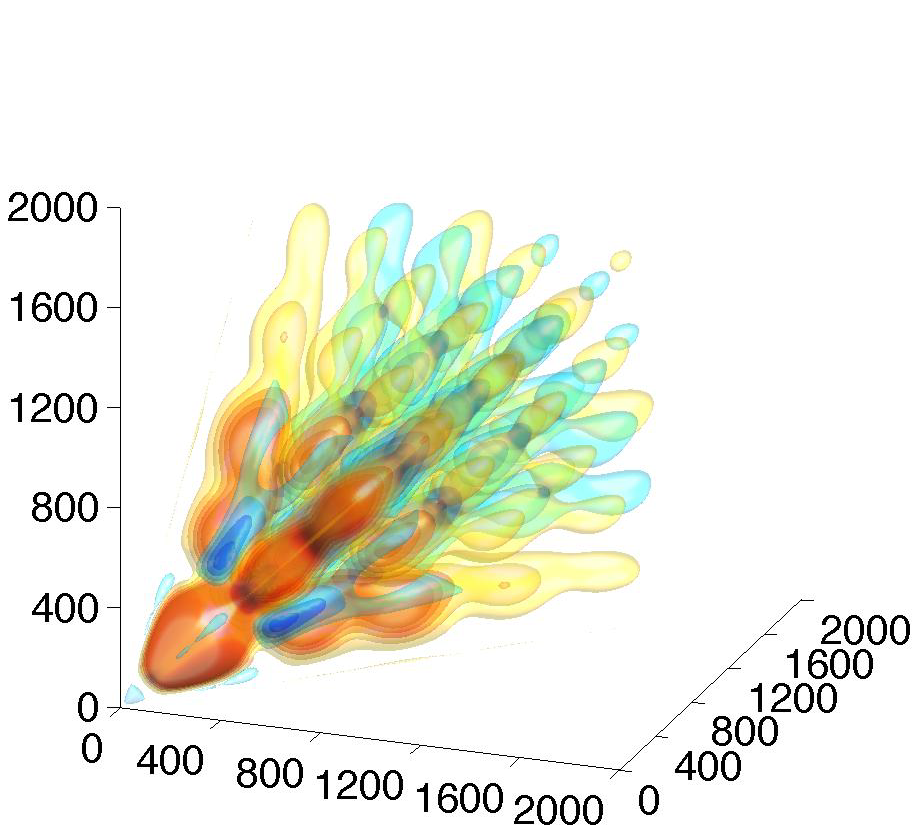} 
\includegraphics[width=0.32\textwidth]{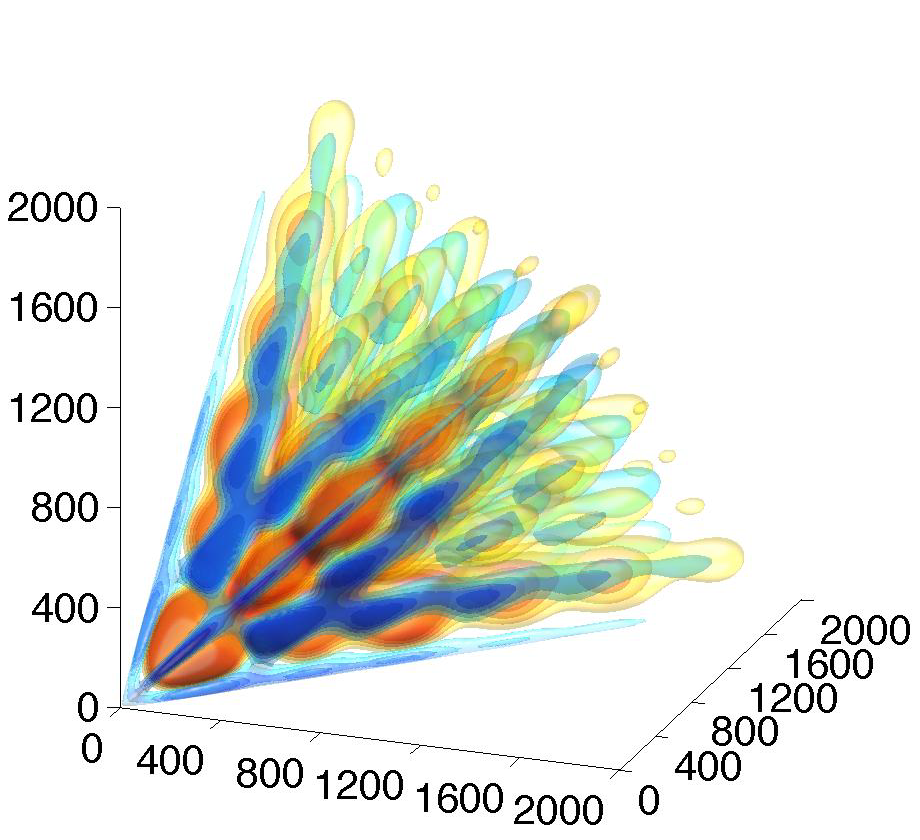} 
\end{center}
\caption{Theoretical predictions for the reduced bispectrum of the CMB, with inflationary models involving non-Gaussianities of local (left), equilateral (center) and orthogonal (right) type. Figure from \cite{Fergusson2009,Fergusson2010}.\label{fig:bispectrum_fNL}}
\end{figure}

Figure \ref{fig:bispectrum_fNL} shows the theoretical reduced bispectrum expected from different shapes of primordial non-Gaussianity.

The most studied (and the most constrained) type of non-Gaussianity is the local model \cite{Salopek1990,Gangui1994,Verde2000,Komatsu2001}, for which the contributions to the primordial bispectrum by squeezed triangles ($k_1 \ll k_2 \simeq k_3$) are dominant. In particular, it can describe the non-linear effects of a super-horizon large-scale mode $k_1$ on smaller scales $k_2$ and $k_3$, still sub-horizon at that moment during inflation. In the limit of weak coupling, the potential is fully described locally, and can be split into two components: a linear contribution, coming from a GRF $\Phi_\mathrm{G}$, and a small non-Gaussian term\footnote{This formula can be seen as the truncation at order two of an expansion of $\Phi(\textbf{x})$ in powers of $\Phi_{\mathrm{G}}(\textbf{x})$. The second term describes an initial skewness of the perturbations. The third term would involve the ``second non-linearity parameter", $g_\mathrm{NL}$, and the cube of $\Phi_{\mathrm{G}}(\textbf{x})$; it describes local initial conditions with kurtosis but no skewness. For a more complete treatment see \cite{Salopek1990,Gangui1994,Komatsu2001}.},

\begin{equation}
\Phi(\textbf{x})=\Phi_{\mathrm{G}}(\textbf{x})+f_{\mathrm{NL}}^{\mathrm{local}} \left( \Phi^{2}_{\mathrm{G}}(\textbf{x}) - \left< \Phi_{\mathrm{G}}^2(\textbf{x}) \right> \right) .
\label{eq:localfNL}
\end{equation}

\begin{figure}
\begin{center}
\begin{overpic}[width=\textwidth]{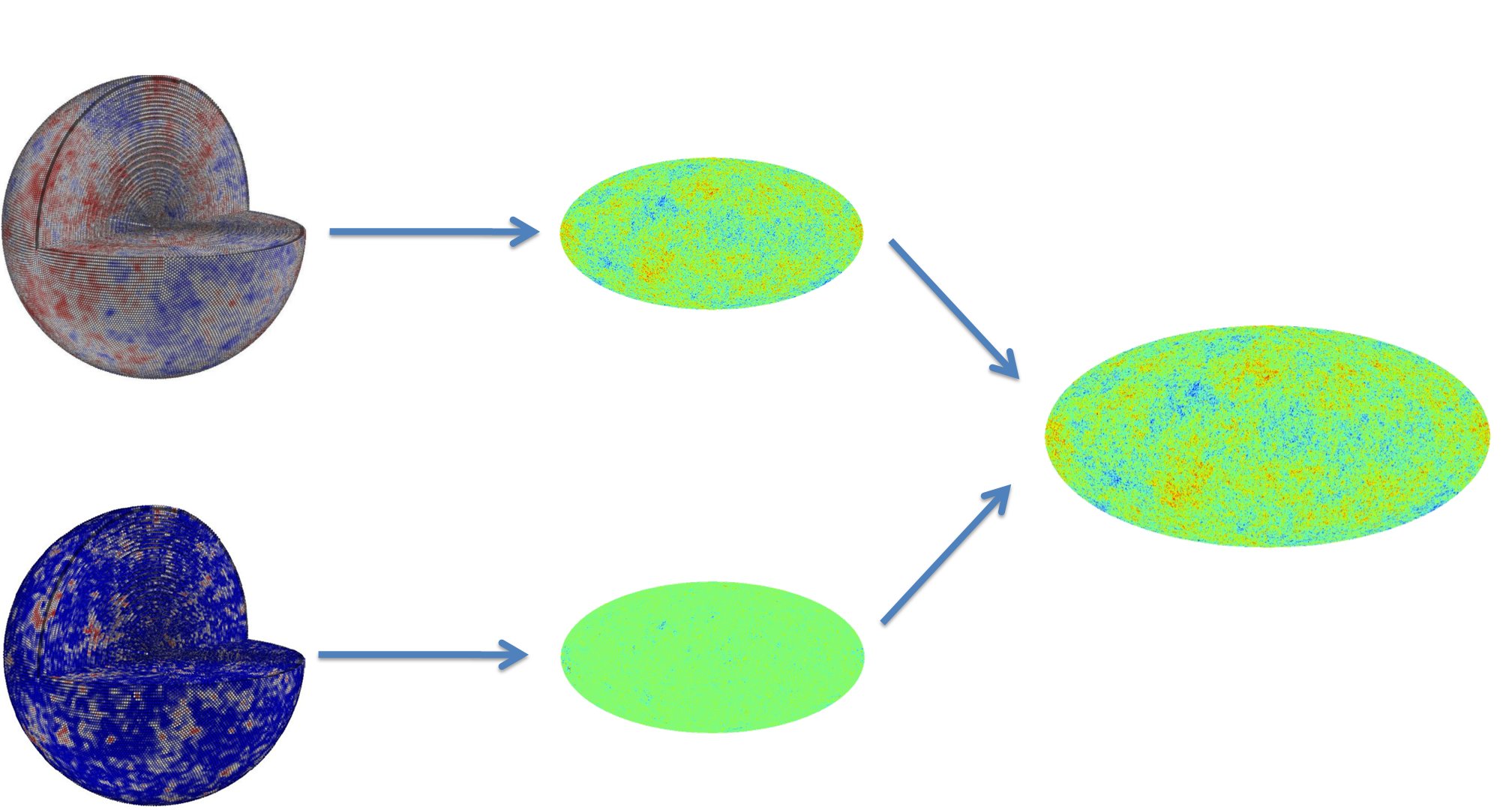}
\put(16,194){Gaussian}
\put(11,85){Non-Gaussian}
\put(308,132){Signal}
\end{overpic}
\end{center}
\caption{In the case of local non-Gaussianity, the CMB signal is obtained from the sum of contributions from a Gaussian and a non-Gaussian potential. Figure adapted from \cite{Elsner2009}. \label{fig:fNLlocal_curvature_to_CMB}}
\end{figure}

\begin{figure}
\begin{center}
\begin{tabular}{cc}
$f_\mathrm{NL}^{\mathrm{local}} = 0$ & $f_\mathrm{NL}^{\mathrm{local}} = 10^2$ \\
\includegraphics[width=0.48\textwidth]{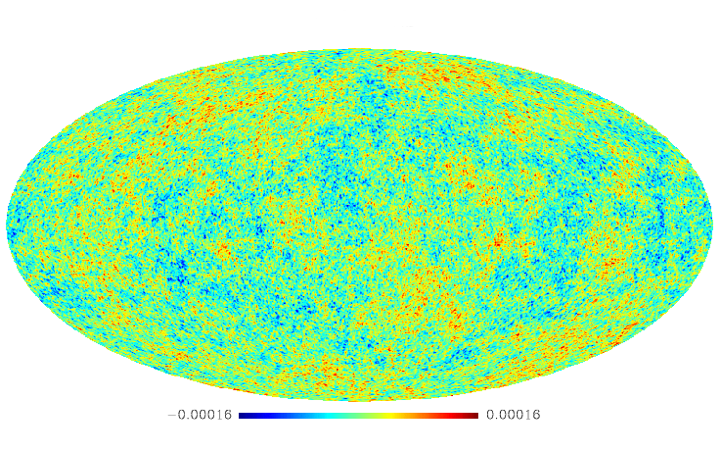} & \includegraphics[width=0.48\textwidth]{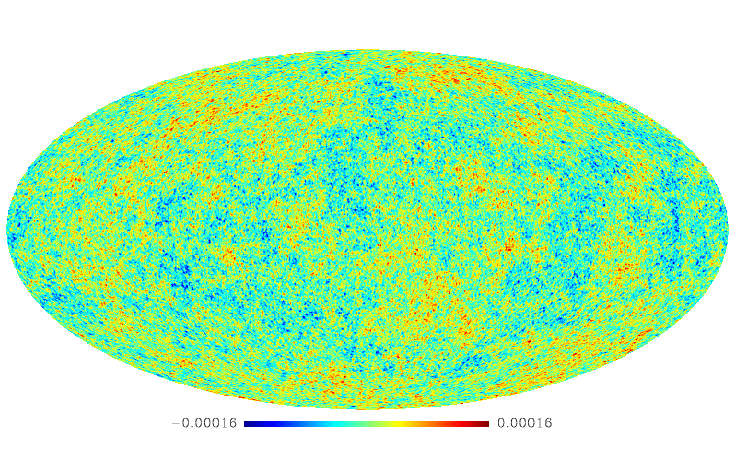} \\ 
$f_\mathrm{NL}^{\mathrm{local}} = 10^3$ & $f_\mathrm{NL}^{\mathrm{local}} = 10^4$ \\ 
\includegraphics[width=0.48\textwidth]{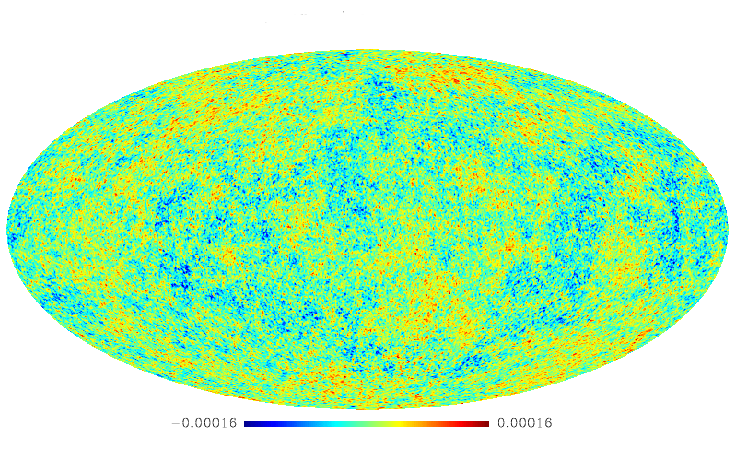} & \includegraphics[width=0.48\textwidth]{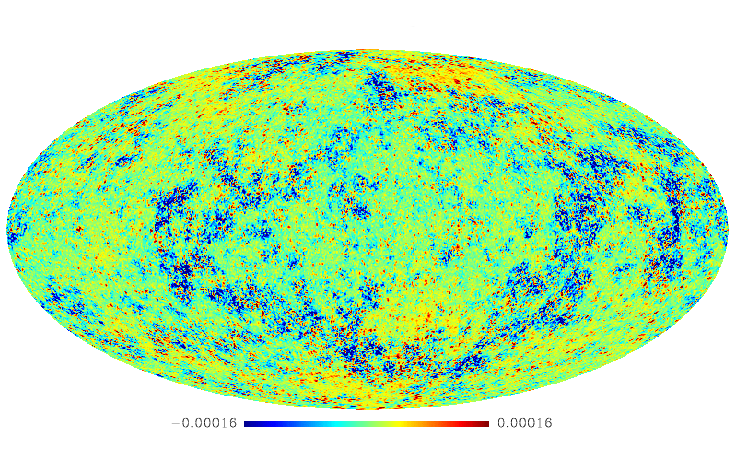}
\end{tabular} 
\end{center}
\caption{Simulated CMB temperature maps with different values of $f_\mathrm{NL}^{\mathrm{local}}$, as indicated above the maps. Figure from \cite{Yadav2005}. \label{fig:fNLlocal_CMBmaps}}
\end{figure}

In this case, the CMB signal would be a mixture of Gaussian and non-Gaussian maps, as illustrated in figure \ref{fig:fNLlocal_curvature_to_CMB}. In figure \ref{fig:fNLlocal_CMBmaps}, we show the effect of different values of $f_\mathrm{NL}^{\mathrm{local}}$ on the CMB temperature map. 

\begin{figure}
\begin{center}
\begin{tabular}{cccc}
\small $\Phi_\mathrm{long}$ & \small $\Phi_\mathrm{short}$ & \small $\Phi_\mathrm{G} = \Phi_\mathrm{long} + \Phi_\mathrm{short}$ & \small $\Phi_\mathrm{NG} = \Phi_\mathrm{G} + f_\mathrm{NL}^{\mathrm{local}} \, \Phi_\mathrm{G}^2$ \\
\includegraphics[width=0.22\textwidth]{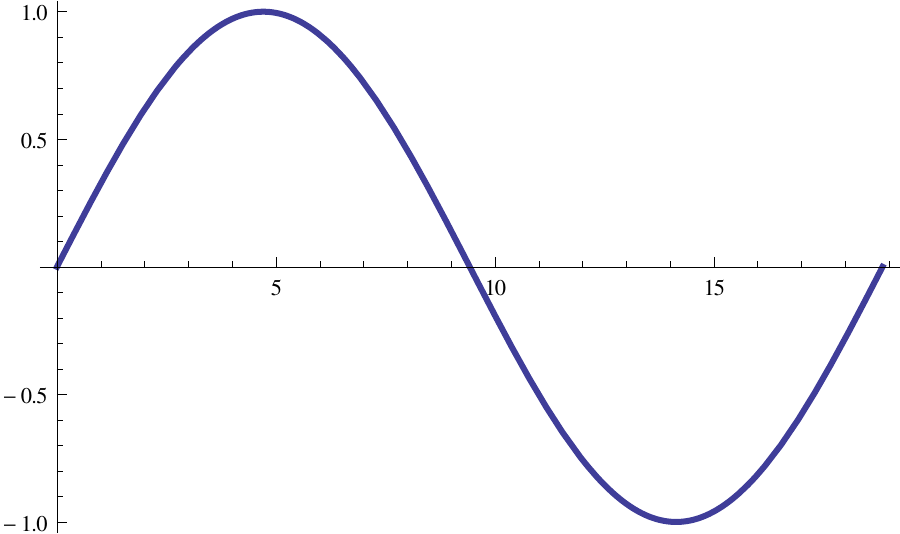} & 
\includegraphics[width=0.22\textwidth]{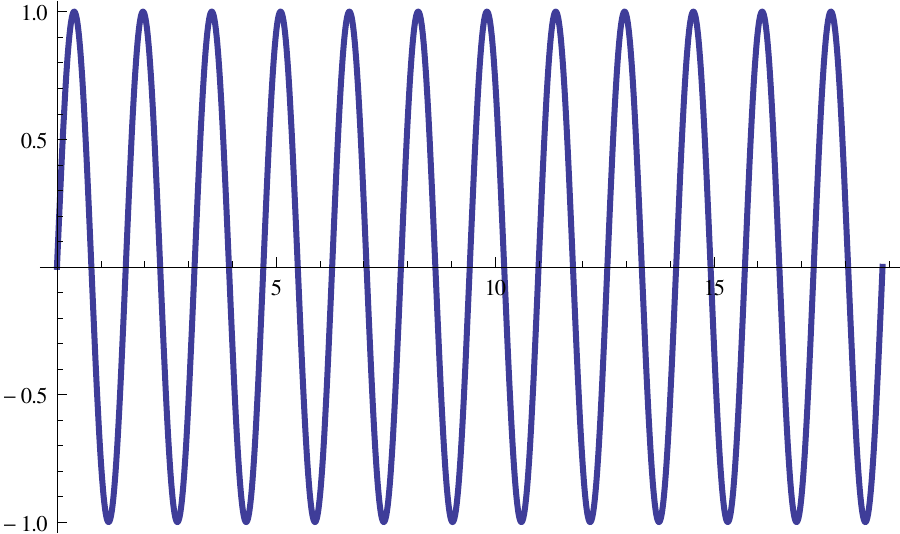} & 
\includegraphics[width=0.22\textwidth]{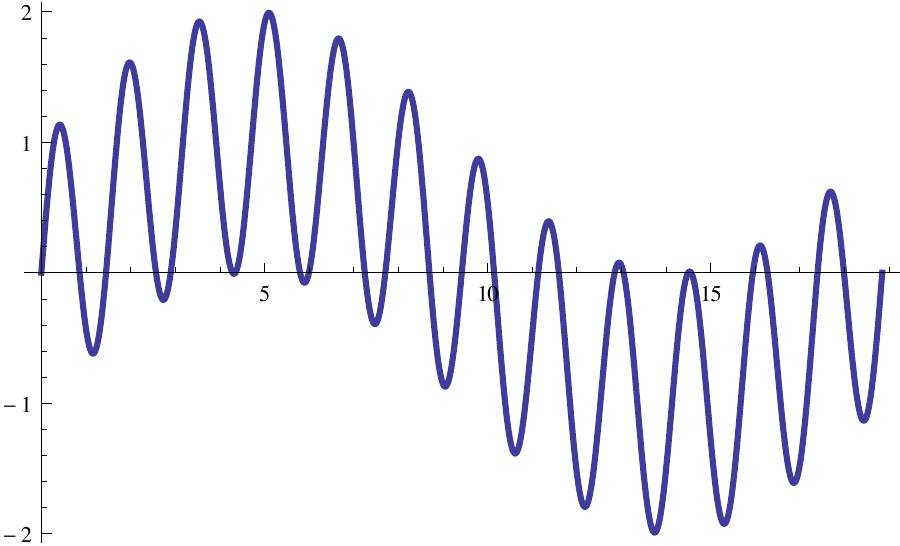} & 
\includegraphics[width=0.22\textwidth]{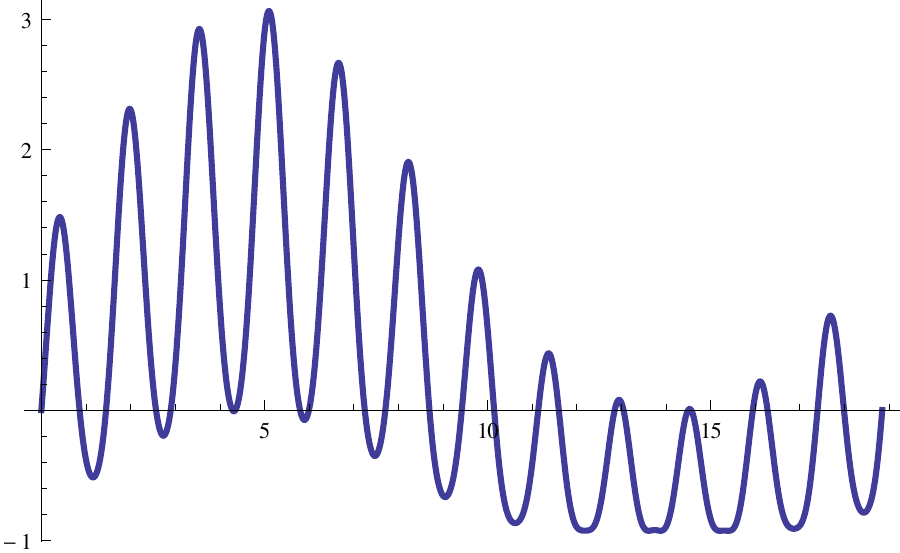}
\end{tabular} 
\end{center}
\caption{Illustration of local non-Gaussianity in Fourier space. The linear contribution $\Phi_\mathrm{G}$ (third panel) is the sum of a long ($\Phi_\mathrm{long}$, first panel) and a short ($\Phi_\mathrm{short}$, second panel) wavelength contribution. The total potential $\Phi_\mathrm{NG} = \Phi_\mathrm{long} + \Phi_\mathrm{short} + f_\mathrm{NL}^{\mathrm{local}} (\Phi_\mathrm{long} + \Phi_\mathrm{short})^2$ (see eq. \eqref{eq:localfNL}) exhibits non-Gaussian features. Coupling of long and short wavelength modes enhances structure in regions of high potential (cold spots) and smoothes structure in regions of low potential (hot spots), as can also be seen in fig. \ref{fig:fNLlocal_CMBmaps}.\label{fig:modes}}
\end{figure}

In Fourier space, it is convenient to distinguish between short (``peaks") and long (the ``background") wavelength modes. This is the peak-background split model, first analysed by \cite{Kaiser1984} in the context of the clustering of galaxies. While these modes are not coupled for a GRF, local non-Gaussianity couples short and long wavelength modes. A value of $f^{\mathrm{local}}_{\mathrm{NL}} > 0$ would enhance structures in regions of high potential (cold spots) and smooth structures in regions of low potential (hot spots) (see figure \ref{fig:modes}).

Primordial non-Gaussianity inference from the CMB map directly sheds light on inflation. Unfortunately, the analysis meets several challenges:

\begin{itemize}
\item the presence of foregrounds\footnote{See \cite{PlanckCollaboration2013ComponentSeparation} for the detailed analysis. In section \ref{sec:The late-time bispectrum}, we discuss the late-time bispectrum that has to be subtracted for primordial NG inference.},
\item systematic errors,
\item computational requirements for the analysis\footnote{See section \ref{sec:Primordial non-Gaussianity inference}.},
\item human cognitive biases\footnote{See section \ref{sec:Primordial non-Gaussianity inference}.}.
\end{itemize}

\subsubsection{The late-time bispectrum}
\label{sec:The late-time bispectrum}

-- Before discussing primordial non-Gaussianity, we emphasize that non-Gaussian features of the CMB also probe late-time fundamental physics. Generally, a time-varying gravitational potential affects the photons on the line of sight. In particular, the erasure of structure caused by the dynamical effect of dark energy (i.e. by the late-time accelerated expansion) causes a late-time decay of the gravitational potential. This phenomenon -- in the linear regime -- is known as the Integrated Sachs-Wolfe (ISW) effect \cite{Sachs1967}. The non-linear growth of density fluctuations also modifies the gravitational potential felt by photons. This is known as the Rees-Sciama effect \cite{Rees1968}. Finally, the weak gravitational lensing of the CMB by the inhomogeneous matter distribution on the line of sight re-maps the temperature primary anisotropy.

\begin{figure}
\begin{center}
\includegraphics[width=0.32\textwidth]{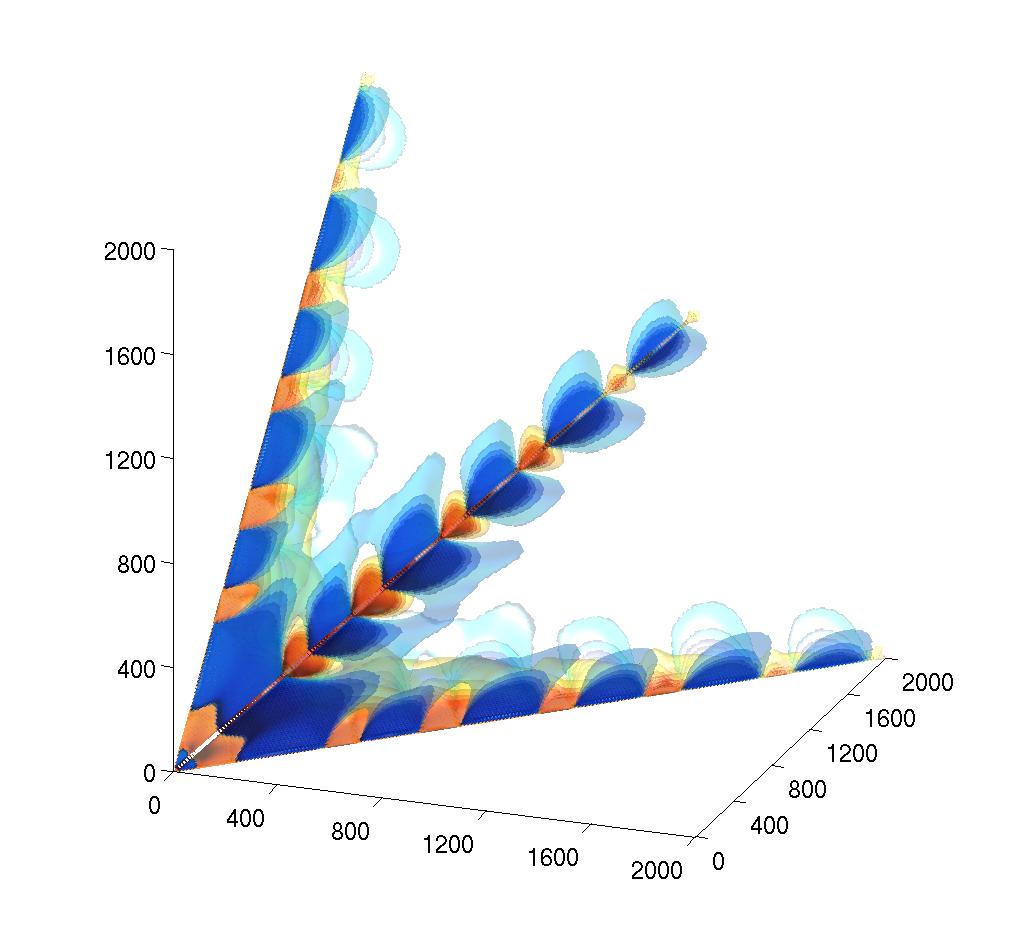}
\includegraphics[width=0.32\textwidth]{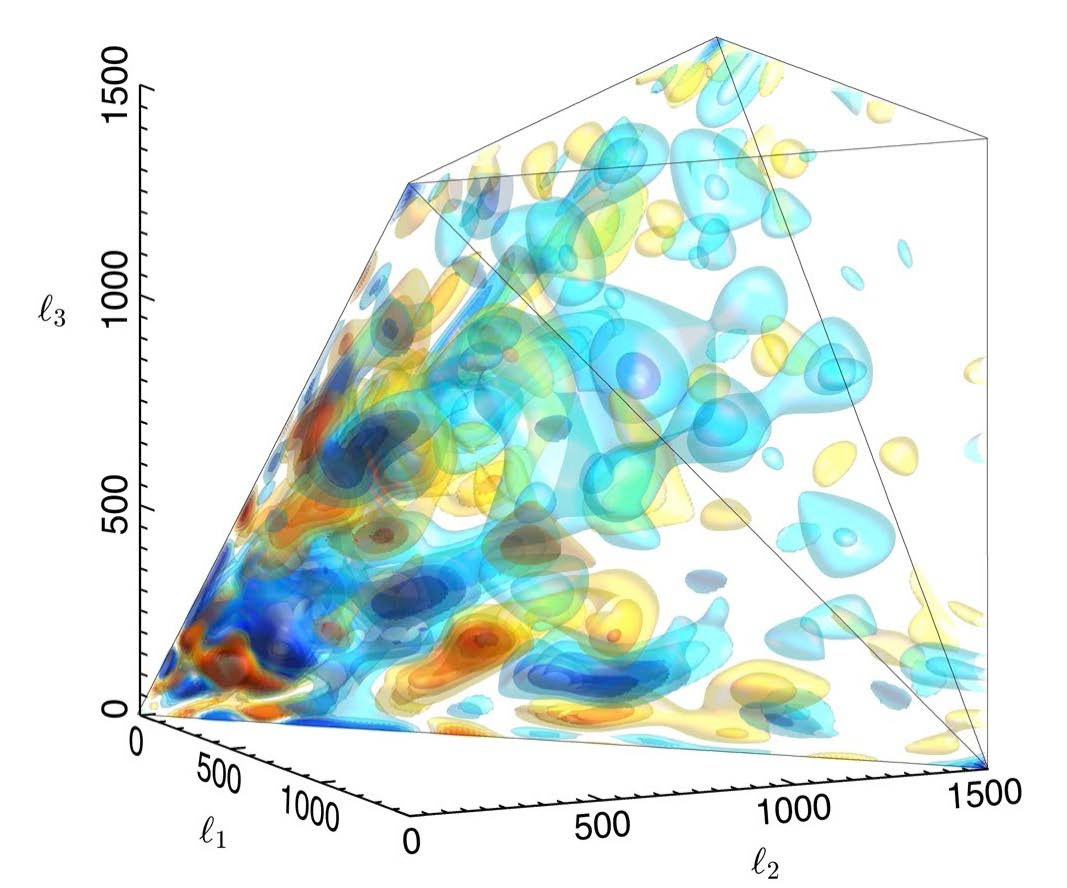}
\includegraphics[width=0.32\textwidth]{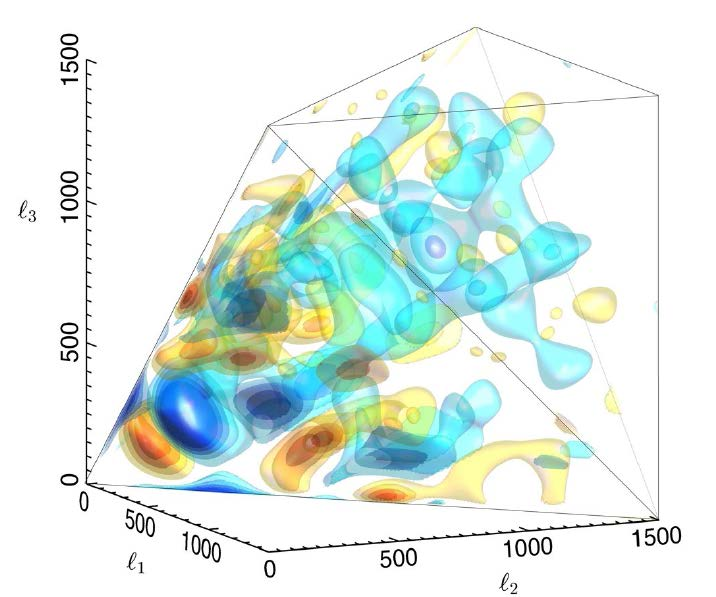}
\end{center}
\caption{\textit{Left panel}. Expected reduced bispectrum signal for the ISW-lensing effect. \textit{Central panel}. Reduced bispectrum estimated by modal decomposition in the Planck SMICA temperature map. \textit{Right panel}. Same reduced bispectrum, after subtraction of the ISW lensing bispectrum template. Note that hints of the ISW lensing effect are visible by eye in the Planck bispectrum. Figure from \cite{PlanckCollaboration2013NonGaussianity}, left panel courtesy of Fergusson and Shellard.\label{fig:ISW_bispectrum}}
\end{figure}

These effects create a non-Gaussian signal in the CMB known as the ``ISW-lensing" bispectrum (fig. \ref{fig:ISW_bispectrum}, see also \cite{Mangilli2013} for an optimal estimator). Planck finds evidence for this effect at the expected level with a $2.6 \, \sigma$ significance \cite{PlanckCollaboration2013ISW}.

\subsubsection{Primordial non-Gaussianity inference}
\label{sec:Primordial non-Gaussianity inference}

-- In this section we discuss functional aspects of primordial non-Gaussianity inference from the Planck CMB map: computational requirements, validation procedures for the analysis and human biases.

This inference problem is a natural playground for the application of Bayesian non-linear techniques. In particular, Bayesian model comparison yields the relative probability of non-zero $f_\mathrm{NL}$ and zero $f_\mathrm{NL}$ given the data. In \cite{Elsner2010b}, the implementation of a fully probabilistic model using Hamiltonian Markov Chain Monte Carlo is presented. Application to Gaussian and non-Gaussian simulations of Bayesian and frequentist methods yield consistent estimates of $f_\mathrm{NL}$. However, sampling the posterior distribution is computationally much more expensive than using a frequentist estimator.

For weak non-Gaussianity, the bispectrum contains all the information about $f_\mathrm{NL}$ \cite{Babich2005} and the bispectrum likelihood for $f_\mathrm{NL}$ is Gaussian \cite{Smith2011}. In this limit, the Bayesian analysis becomes therefore equivalent to determining value and error bar of the bispectrum estimator, which is computationally cheaper. The Planck analysis uses this simplification. Furthermore, the brute force evaluation of the usual estimator, $\hat{f}_\mathrm{NL}$ (see e.g. section 3.5 in \cite{Liguori2010}), scaling as $O(\ell_\mathrm{max}^5)$, is unfeasible for Planck data. The key idea to get around this issue is the Komatsu-Spergel-Wandelt (KSW) factorization \cite{Komatsu2005}, which yields a $O(\ell_\mathrm{max}^2)$ speed up, corresponding to a factor of about $\sim 10^6$ - $ 10^{7}$ for Planck.

Different bispectrum estimators exist and have been used on the Planck data:
\begin{itemize}
\item KSW \cite{Komatsu2005} gives an exact fit to separable bispectra (local, equilateral and orthogonal), but is limited to factorizable templates.
\item Modal \cite{Fergusson2010} is a highly efficient generalization that uses sums of KSW-like smooth templates to fit arbitrary templates, allowing to look for more general primordial models. It fits arbitrary bispectrum template up to resolution limit, not just the separable kinds.
\item Binned \cite{Bucher2010} uses block-shaped templates, binning in different $\ell$-ranges. Data compression of the smoothed observed bispectrum allows to reduce the computational weight of the analysis.
\item Skew-$C_\ell$ extension \cite{Munshi2010}.
\item Minkowski functionals \cite{Ducout2013}.
\end{itemize}

To participate in the analysis, all estimators have to pass a suite of validations on Gaussian and non-Gaussian simulations. These simulations test map-to-map agreement between estimators and robustness to foreground residuals. A blind challenge of recovering unknown non-Gaussianity from a realistically simulated map (including foregrounds, correlated noise and realistic systematic uncertainty) has also been part of the validation procedure.

The Planck NG analysis passed an extensive validation campaign. Using different bispectrum estimators, consistent values for the primordial local, equilateral, and orthogonal bispectrum amplitudes have been obtained. The results have been demonstrated to be stable for the four different foreground cleaning methods (SMICA, NILC, SEVEM, and C-R), with negligible impact of foreground residuals. They have also passed a suite of tests studying the dependence on resolution ($\ell_\mathrm{max}$), on frequency channels and on the mask, as well as several null tests.

Another verification used for Planck data is a check for consistency with the previous CMB anisotropy experiment, WMAP. Limiting the analysis to the same multipole moment as the WMAP resolution, the Planck analysis finds results consistent with WMAP9 values \cite{Bennett2013} ($f_\mathrm{NL}^{\mathrm{local}} = 37.2 \pm 19.9$). With the use of the whole Planck data set (including ten times more modes), the central WMAP9 value is ruled out by $\sim 6 \, \sigma$, illustrating the extreme precision of Planck.

Before we quote the final Planck results on primordial non-Gaussianity, we pause and think about the consequences they may have on current research in cosmology. A robust detection of $f_\mathrm{NL}$ would rule out standard inflation and mean a paradigm-shift in our understanding of the early universe. For such a revolution to happen, the conclusion would have to be extremely solid. Conversely, Gaussian initial conditions could mean that the universe is less complicated -- at least at first sight -- which might be considered reassuring. In any case, we have to be very careful to produce an unbiased analysis, in particular free from human biases. 

Since the method heavily relies on statistical inferences, the problem of multiple comparisons (the ``look-elsewhere" effect) can show up: apparently statistically significant observations may arise by chance, just because of the size of the parameter space to be searched. This means that incorrect rejection of the null hypothesis is more likely to occur when one considers multiple inferences from a large data set. The usual way to escape this bias is to define \textit{a priori} the hypotheses whose statistical significance have to be tested and to avoid \textit{a posteriori} predictions. If this is not possible, one should require a stronger level of evidence for an individual effect to be deemed ``significant", so as to compensate for the number of inferences being made.

Another possible bias in scientific investigation is the tendency to favor information that confirms one's beliefs or hypotheses (e.g. a preference for a Gaussian versus a non-Gaussian universe). This ``confirmation bias" is displayed when we interpret data toward confirming our existing beliefs or when we test ideas in a one-sided way, focusing on one possibility and ignoring alternatives.

\begin{figure}
\begin{center}
\includegraphics[width=0.6\textwidth]{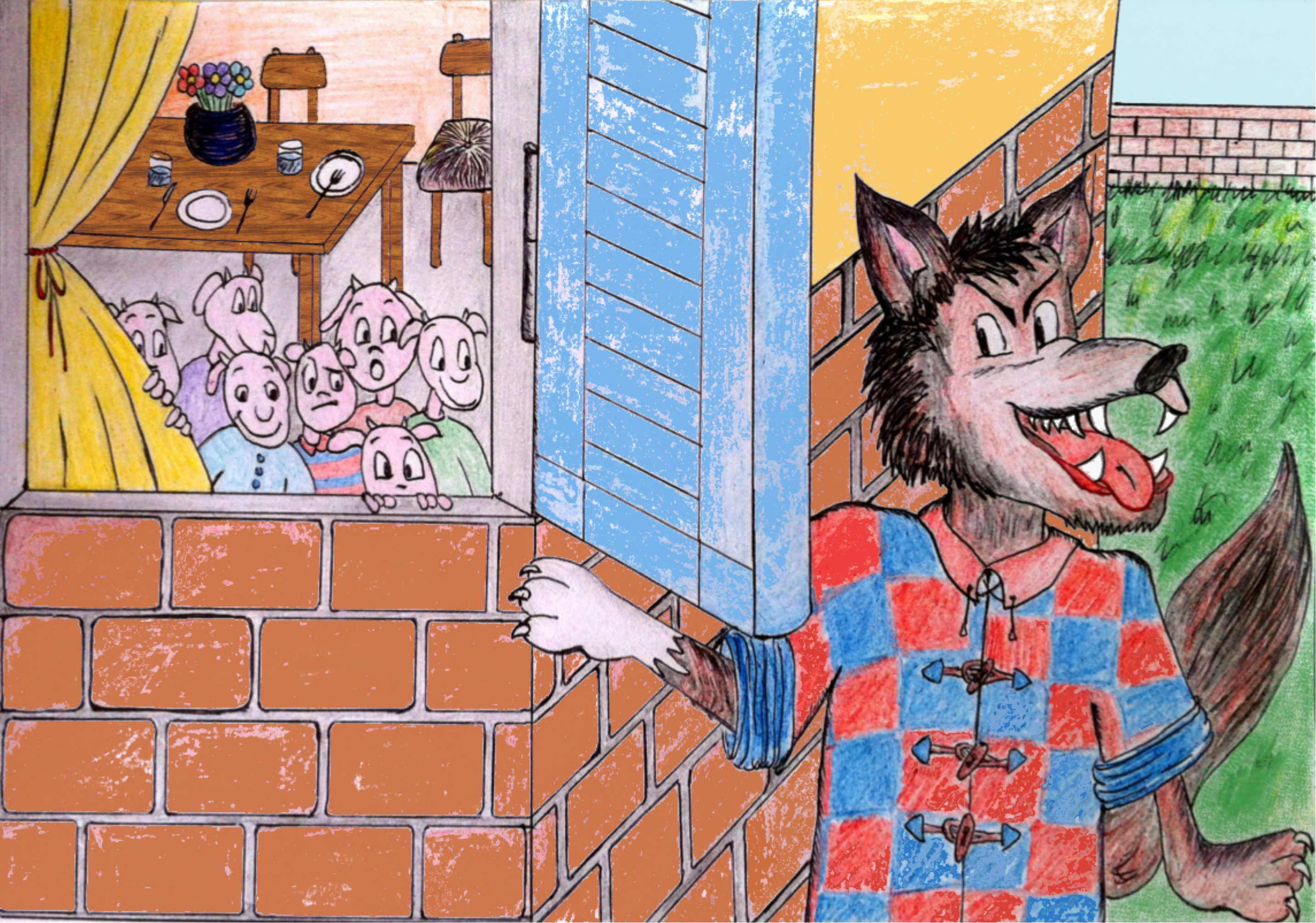} 
\end{center}
\caption{``The wolf and the seven little kids": a tale of confirmation bias.\label{fig:wolf}}
\end{figure}

The tale ``The wolf and the seven little kids" \cite{Grimm1812} (fig. \ref{fig:wolf}) is a story of confirmation bias. The kids are waiting for their mom to come home -- they want to ``detect" their mom. At first the wolf's voice does not match the mother's. The first-look test fails and the kids are safe. Then the wolf changes his voice. By chance, the kids require a systematic test: ``Show us your paw!". The test fails and the kids are safe. When the wolf paints his paw white, the test passes. Instead of doing further tests (``Show us your face!", ``Show us your tail!"), the kids let the wolf in, and he eats them all up.

In cosmology, we have a strong prior: a canonical Gaussian $\Lambda$CDM universe. Red flags rise in the analysis if we depart from this model. In order to ``avoid being eaten", all measurements of primordial non-Gaussianity should be treated on an equal footing, for example undergo the same battery of tests, regardless of whether it is a detection or a constraint. The care to avoid human biases also was at the basis for the extensive validation campaign of results discussed before, involving different estimators and component separation techniques.

\subsubsection{Planck constraints on inflationary models}

-- The final results quoted for primordial non-Gaussianity (measurement on the SMICA map with the optimal KSW estimator, after subtraction of the ISW-lensing contribution) are the following, at 1-$\sigma$ confidence level:

\begin{eqnarray}
f_\mathrm{NL}^{\mathrm{local}} & = & 2.7 \pm 5.8\\
f_\mathrm{NL}^{\mathrm{equil}} & = & −42 \pm 75\\
f_\mathrm{NL}^{\mathrm{ortho}} & = & −25 \pm 39
\end{eqnarray}

As we have discussed, many tests validate the robustness of this result. There is no evidence for primordial NG of one of these shapes. The new constraint volume for the three main types of NG is 20 times smaller than before Planck, making of these constraints the highest precision test to date of physical mechanisms for the origin of cosmic structure. The view of the initial state of the universe after Planck supports the simplest models: slowly-rolling, single scalar-field inflationary models are favoured. All results are consistent with a non-excited Bunch-Davies initial vacuum state. Multi-field models are not ruled out, but are also not detected. In particular, the curvaton decay fraction $r_\mathrm{D} \equiv [3\rho_\mathrm{curvaton}/(3\rho_\mathrm{curvaton}+4\rho_\mathrm{radiation})]_\mathrm{D}$ (evaluated at the epoch of the curvaton decay) is constrained to be $\geq 15\%$ (95\% CL). Planck rules out small speed of sound during inflation: $c_\mathrm{s} \geq 0.02$ (95\% CL) and strongly constrains models such as DBI, $k$-inflation and warm inflation. Finally, the data put severe pressure on a class of ekpyrotic/cyclic scenarii (those with exponential potential, entropic generation of perturbations and conversion during the ekpyrotic smoothing phase).

As a conclusion, with these results, the paradigm of standard single-field slow-roll inflation has survived its most stringent tests to date.
\section*{Conclusion: From theory to data -- There and Back Again... A cosmologist's tale}

\begin{figure}
\begin{center}
\includegraphics[width=0.5\textwidth]{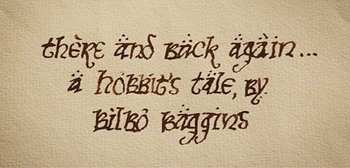}
\caption{The reference behind our conclusion's title \cite{Baggins_TA_3001}\label{fig:ThereAndBackAgain}.}
\end{center}
\end{figure}

As we have shown repeatedly, having either just data or just theory is not enough to create meaningful inferences. Data generates and shapes new theoretical insights, just as theory is necessary to lend support to data analysis (fig. \ref{fig:ThereAndBackAgain}). Theory, when rooted in theoretical consistency and elegance, can anticipate the experimental results that will guide our investigations. Then, as data come in, theorists will be forced to develop, to revisit or to abandon some suggested explanations. Even if hypotheses are initially well-motivated, it is ultimately observations that determine what is correct.

In this framework it is of utmost importance to design experiments and surveys such as to target current uncertainty in the theory. Evidence identification and accumulation procedures also hold crucial implications for the design of future assessments, and thus have implications for the development of conceptual models. In the next few years, new cosmological insights will come from ongoing, upcoming or proposed experiments. Among them, high-redshift galaxy surveys (\textsc{Hetdex}\footnote{http://hetdex.org/}, \textsc{SDSS IV: eBOSS}\footnote{http://www.sdss3.org/future/eboss.php}, \textsc{DESI}\footnote{http://desi.lbl.gov/}, \textsc{LSST}\footnote{http://www.lsst.org/}, \textsc{Euclid}\footnote{http://sci.esa.int/euclid/}), study of the intensity, polarization, and frequency spectrum of the microwave sky (\textsc{Prism}\footnote{http://www.prism-mission.org/}), of the 21 cm ray and the radio sky (\textsc{SKA}\footnote{http://www.skatelescope.org/}), and of the gravitational waves sky (\textsc{eLISA}\footnote{https://www.elisascience.org/}).

Surprising or unexpected data may or may not show up, but regardless of what happens, the interplay between theory and data will lead us to expand our knowledge and to make progress in our interpretation of the Universe.

\acknowledgments
BDW would like to express his gratitude to the organizers of the Enrico Fermi school for the opportunity to deliver these lectures and to the graduate students of the Paris \'Ecole Doctorale 127 for Astronomy and Astrophysics for their feedback and participation in his courses related to this topic. It was a real pleasure for the authors to stay in Varenna and benefit from inspiring discussions in such beautiful environment and we thank the organizers for their generous hospitality. Any Planck results referred to in these lectures were presented by BDW on behalf of the Planck collaboration. We are grateful to Jens Jasche for useful comments on the draft version. FL acknowledges funding from an AMX grant (\'Ecole polytechnique ParisTech). AP and BDW acknowledge support from BDW's Chaire Internationale in Theoretical Cosmology at the Universit\'{e} Pierre et Marie Curie. BDW gratefullly acknowledges a senior Excellence Chair of the Agence Nationale de la Recherche (ANR-10-CEXC-004-01),   NSF grants AST 07-08849, NSF AST 09-08693 ARRA and NSF AST 09-08902, and NASA subcontract JPL 1236748. This work made in the ILP LABEX (under reference ANR-10-LABX-63) was supported by French state funds managed by the ANR within the Investissements d'Avenir programme under reference ANR-11-IDEX-0004-02.

\bibliographystyle{varenna}
\bibliography{Varenna_proceedings}

%
\end{document}